\documentclass{article}

\usepackage[preprint]{neurips_2026}

\usepackage[utf8]{inputenc}
\usepackage[T1]{fontenc}
\usepackage{hyperref}
\usepackage{url}
\usepackage{booktabs}
\usepackage{amsfonts}
\usepackage{amsmath}
\usepackage{amssymb}
\usepackage{nicefrac}
\usepackage{microtype}
\usepackage{wrapfig}
\usepackage{xcolor}
\usepackage{graphicx}
\graphicspath{{figs/}}
\usepackage{enumitem}
\usepackage[normalem]{ulem}
\setlist[itemize]{topsep=2pt,partopsep=0pt,itemsep=1pt,parsep=0pt}
\setlist[enumerate]{topsep=2pt,partopsep=0pt,itemsep=1pt,parsep=0pt}
\makeatletter
\renewcommand{\paragraph}{%
  \@startsection{paragraph}{4}{\z@}%
                {0.6ex \@plus 0.3ex \@minus 0.2ex}%
                {-1em}%
                {\normalsize\bf}%
}
\makeatother

\title{AudioCALM: Continuous Autoregressive Language Modeling for Universal Audio Generation}

\author{
  \textbf{Huadai Liu}$^{1,2}$, \textbf{Kaicheng Luo}$^{2}$, \textbf{Wen Wang}$^{2}$, \textbf{Qian Chen}$^{2}$ \\
   \textbf{Bin Ma}$^{2}$, \textbf{Xiangang Li}$^{2}$, \textbf{Wei Xue}$^{1}$~\thanks{Corresponding author.} \\
  $^{1}$Hong Kong University of Science and Technology (HKUST) \\
  $^{2}$Tongyi Fun Team, Alibaba Group \quad
}

\begin{document}

\maketitle

\begin{abstract}
Unifying speech, sound, and music generation in one model is hindered by tradeoffs between fidelity, end-to-end training, in-context conditioning, and variable-length synthesis that no current paradigm fully resolves. To address this challenge, we present \textbf{AudioCALM}, a universal audio generation framework that extends autoregressive (AR) next-token prediction from discrete tokens to continuous audio latents: a thin flow-matching head replaces the softmax to predict rectified-flow velocities at each position, and a block-causal AR-Flow attention pattern produces arbitrary-length output. Joint training of multiple audio generation tasks faces an asymmetric text--audio mismatch: speech transcripts align to specific time spans and demand \textit{tight, time-aligned} attention, whereas sound and music captions describe only overall semantics and rely on \textit{diffuse, holistic} attention; mixing the two disproportionately degrades sound and music generation. We address this asymmetry at two levels: a \textbf{data reformulation} strategy that unifies all three tasks under a single description-style conditioning interface, and a novel architecture \textbf{Asymmetric Mixture-of-Modality-Experts (A-MoME)}, which adds a dedicated residual expert for speech while sound and music share the backbone, incurring no inference overhead on non-speech inputs. Experimental results demonstrate that AudioCALM matches modality-specific state-of-the-art and outperforms prior unified baselines on speech, sound, and music generation benchmarks. The project page is at \url{https://AudioCALM-Project.github.io}.
\end{abstract}

\vspace{-1mm}
\section{Introduction}
\label{sec:intro}

One long-standing goal of audio generation is building a universal model that could synthesize a wide range of audio types (that is, modalities or domains), such as speech, music, and environmental sound,  from natural-language conditioning in a unified manner: real-world audio rarely occurs in isolation, and a unified model can in principle leverage data and shared structure and knowledge across various audio domains, hence reducing development cost and improving generation performance. In practice, however, the field remains dominated by domain-specific systems built on three generative paradigms: discrete-token autoregressive (AR) modeling over neural audio codecs~\citep{valle, musicgen, audiogen, audiolm}, cascaded language-model-then-acoustic-generator pipelines~\citep{cosyvoice2, maskgct, naturalspeech3}, and non-autoregressive (NAR) latent diffusion or flow matching over continuous representations~\citep{audioldm2, stableaudioopen, voicebox, matchatts}. Each of these paradigms imposes structural constraints on its generative interface, limiting how uniformly a single instantiation can support text-to-speech (TTS), text-to-sound, and text-to-music (T2M) tasks. Note that ``Audio'' is our umbrella term for speech, sound, and music, and T2A denotes the text-to-non-speech regime (sound and music together). On the other hand, truly unified attempts spanning all three modalities are far fewer, concentrated on autoregressive language modeling over discrete audio tokens~\citep{uniaudio, uniaudio15} and non-autoregressive flow matching with task-specific input formatting~\citep{audiobox,uniflowaudio}, with the cascaded paradigm comparatively less explored under the same unified setting.

However, all three paradigms suffer from severe limitations. Specifically, discrete-token autoregression~\citep{valle, audiogen, audiolm} imposes its constraint by compressing audio into a finite codebook, an information bottleneck especially restrictive for high-bandwidth content such as polyphonic music and environmental sound, where token-based methods trail continuous-latent approaches~\citep{audioldm2, stableaudioopen} in performance. Cascaded systems sidestep discretization by handing off to a separate diffusion or flow-matching module~\citep{cosyvoice2, maskgct}, but the hand-off introduces a rigid intermediate representation and severs end-to-end optimization of the language model and the acoustic generator. Non-autoregressive flow matching models avoid both issues, but forgo the autoregressive in-context conditioning over previously generated content, and depend on external duration prediction or forced alignment for variable-length generation~\citep{naturalspeech3, audiobox}. These limitations motivate four desiderata for universal audio generation---\emph{fidelity}, \emph{end-to-end optimization}, \emph{in-context understanding}, and \emph{natural variable-length generation}---which no existing paradigm fully satisfies.

We address this gap with \textbf{AudioCALM}, a universal audio generation framework instantiating \emph{Continuous Autoregressive Language Modeling (CALM)}: a paradigm that extends autoregressive next-token prediction from discrete tokens to continuous audio latents while retaining the autoregressive language-model backbone. A single LM processes a unified text--audio sequence, and at each audio position, a thin \emph{flow-matching head} on the LM hidden state predicts a rectified-flow velocity over the continuous latent space---the continuous counterpart of the softmax in standard next-token prediction. The two components share one latent space and are jointly trained under a single rectified-flow objective, removing any intermediate discretization. Consequently, different from all three paradigms, \textbf{AudioCALM retains the LM interface of discrete-token autoregressive models, the high-fidelity continuous representation of non-autoregressive flow matching models, and the end-to-end optimization that cascaded LM-then-generator systems sacrifice}. To support streaming arbitrary-length generation, we further introduce \textbf{AR-Flow}, a novel block-causal attention pattern in which each latent block is generated by flow matching while attending autoregressively to all preceding blocks and the full text condition, thus variable length becomes a property of the attention mask rather than a predefined hyperparameter.

Within this unified architecture, text still plays substantially different roles across tasks: TTS exhibits largely \emph{local} text--audio correspondence---transcript segments align to specific time spans---while T2A is largely \emph{global}, with captions fixing only overall semantics. Therefore, speech favors sharp, locally aligned text--audio attention, while sound and music favor diffuse, globally distributed text--audio attention. Our joint-training experiments reveal an asymmetric mismatch (Table~\ref{tab:ablation}): adding speech to the training mixture degrades sound and music generation more than sound and music degrade speech generation, which is consistent with speech's stricter alignment crowding out the dispersed attention that non-speech audio depends on. Symmetric designs that allocate capacity uniformly across modalities, therefore, fail to match the structure of this mismatch.

We address this asymmetry at two levels. At the \emph{data} level, we adopt a single description-style conditioning interface across all three modalities. An audio-conditioned multimodal LLM (MLLM) consumes each training clip together with its bare textual annotation---the transcript for speech, the short caption for sound and music---and emits a long-form description that surfaces modality-relevant attributes from the waveform as natural-language phrases: for speech, speaker timbre, fine-grained prosody, and acoustic context (e.g., reverberation, background noise, recording quality); for sound, acoustic events and their sources, scene, and event sequencing; for music, genre, instrumentation, tempo, mood, and recurring motifs. Under this interface, TTS, text-to-sound, and text-to-music all reduce to conditional next-block prediction over audio latents under one model, and zero-shot voice cloning is supported natively by prepending a speaker embedding as a global prefix. At the \emph{architecture} level, the residual parameter-level mismatch is handled by \textbf{Asymmetric Mixture-of-Modality-Experts (A-MoME)}, an asymmetric variant of modality-expert designs~\citep{vlmo, beit3, mot}: self-attention and the main feed-forward network remain shared across speech, sound, and music, while a speech-specific residual expert is added in parallel to capture the tighter text--audio correspondence required by speech alone. The asymmetry, rather than the modality count, defines the design: sound and music share the backbone unchanged, and only the modality whose objective diverges from the others receives dedicated capacity. Notably, the integration of A-MoME layers into AudioCALM does not modify the flow-matching head or the AR-Flow attention pattern, and incurs no inference overhead on non-speech inputs.

Our contributions are summarized as follows:

\begin{itemize}[leftmargin=*,noitemsep]
    \vspace{-1mm}
    \item We propose \textbf{AudioCALM}, a universal audio generation framework that extends next-token prediction to continuous audio latents via a flow-matching head, paired with a block-causal AR-Flow attention pattern that yields arbitrary-length speech, sound, and music generation in a single model.
    \item We identify an asymmetric local--global text--audio mismatch in joint training and address it at two levels: a data-level recasting of speech as a special case of audio that unifies the three modalities under one description-style conditioning interface, and an architecture-level innovation of \textbf{A-MoME}, which mirrors this asymmetry by adding a dedicated residual expert for speech alone while keeping sound and music to share the backbone, allocating capacity by demand rather than by modality.
    \item AudioCALM achieves state-of-the-art (SoTA) performance on speech, sound, and music generation benchmarks, surpassing both modality-specific systems and prior unified models on per-domain quality across all three audio domains.
\end{itemize}

\vspace{-1mm}
\section{Related Work}
\label{sec:related}

\paragraph{Unified Audio Generation}
We use \emph{unified audio generation} to refer to a single model whose \emph{output modality} jointly covers speech, music, and sound---replacing the historically separate TTS, T2A, and T2M systems---with text as the primary conditioning \emph{input}, optionally augmented with reference-audio conditioning for voice cloning or style imitation. Existing output-unified frameworks fall along three architectural lines.
The first category is discrete-token autoregressive modeling, exemplified by VALL-E~\citep{valle} for TTS, MusicGen~\citep{musicgen} for T2M, AudioGen~\citep{audiogen} for T2A, and the unified UniAudio~\citep{uniaudio, uniaudio15} in the audio language model (AudioLM) family~\citep{audiolm}, which tokenizes audio with a neural codec~\citep{encodec, dac} and casts every task as next-token prediction; this yields a clean LM interface, but the codec bottleneck caps fidelity on polyphonic music and complex sound generation~\citep{audioldm2, stableaudioopen,flashaudio,audiolcm}.
The second category is the cascaded LM-then-generator family, such as CosyVoice2~\citep{cosyvoice2} and MaskGCT~\citep{maskgct}, in which an LM emits semantic tokens for a downstream diffusion or flow-matching synthesizer, removing the fidelity ceiling but freezing the intermediate semantic code as a fixed contract between two separately optimized stages, which blocks end-to-end optimization.
The third category is non-autoregressive flow matching~\citep{thinksound,prismaudio,omniaudio}, e.g., Audiobox~\citep{audiobox}, Stable Audio Open~\citep{stableaudioopen}, and the recent UniFlow-Audio~\citep{uniflowaudio}, which conditions flow-matching networks on task-specific input formatting and attains strong fidelity, but generates in parallel and thereby requires an externally specified target length and drops the token-level in-context conditioning of AR models~\citep{naturalspeech3}.
Orthogonal to the three output-unified lines above, a related strand of speech-output universal models---early speech-only LMs~\citep{speechgpt, audiopalm, moshi} and recent omni-modal audio LMs~\citep{kimiaudio, qwen25omni, mimoaudio, funaudiochat}---ingests speech, music, and sound as \emph{input} for understanding but produces only speech (and text) as \emph{output}, leaving music and sound \emph{generation} unaddressed; concurrent work~\citep{unimoeaudio} tackles the resulting cross-modal interference via mixture-of-experts in a unified speech--music generator, yet still excludes general sound.

\paragraph{Audio Language Models}
Among the three paradigms above, the AR LM family uniquely affords the standard LLM toolbox---streaming inference, in-context conditioning, and scalable training---yet discrete-token AR is bounded by lossy codec reconstruction and intricate multi-codebook residual-quantization decoding patterns, while continuous-latent diffusion and flow matching~\citep{voicebox, rectifiedflow, vitttts} avoid the codec ceiling only by sacrificing the streaming and token-level in-context behavior of AR models.
A more recent line resolves this tension by preserving the LM interface while removing the codec bottleneck: it replaces the discrete softmax with a continuous output head, instantiated by GIVT~\citep{givt} and MAR~\citep{mar} in vision and by DiTAR~\citep{ditar} in audio, where a flow-matching head is attached to LM hidden states for speech generation alone.
AudioCALM extends this continuous-latent AR line from the speech-only DiTAR to universal audio generation. Beyond this domain extension, AudioCALM further differs from DiTAR in three respects: (i) we directly use the LLM itself as the denoiser across speech, sound, and music; (ii) we introduce \textbf{A-MoME} to resolve the cross-modal local--global text--audio mismatch that does not arise in DiTAR's single-modality setting; and (iii) we recast speech as a special case of audio under a unified description-style conditioning interface, so that TTS, T2A, and T2M reduce to one conditional next-block prediction problem within a single model.

\begin{figure}[t]
    \vspace{-2mm}
    \centering
    \includegraphics[width=.95\linewidth]{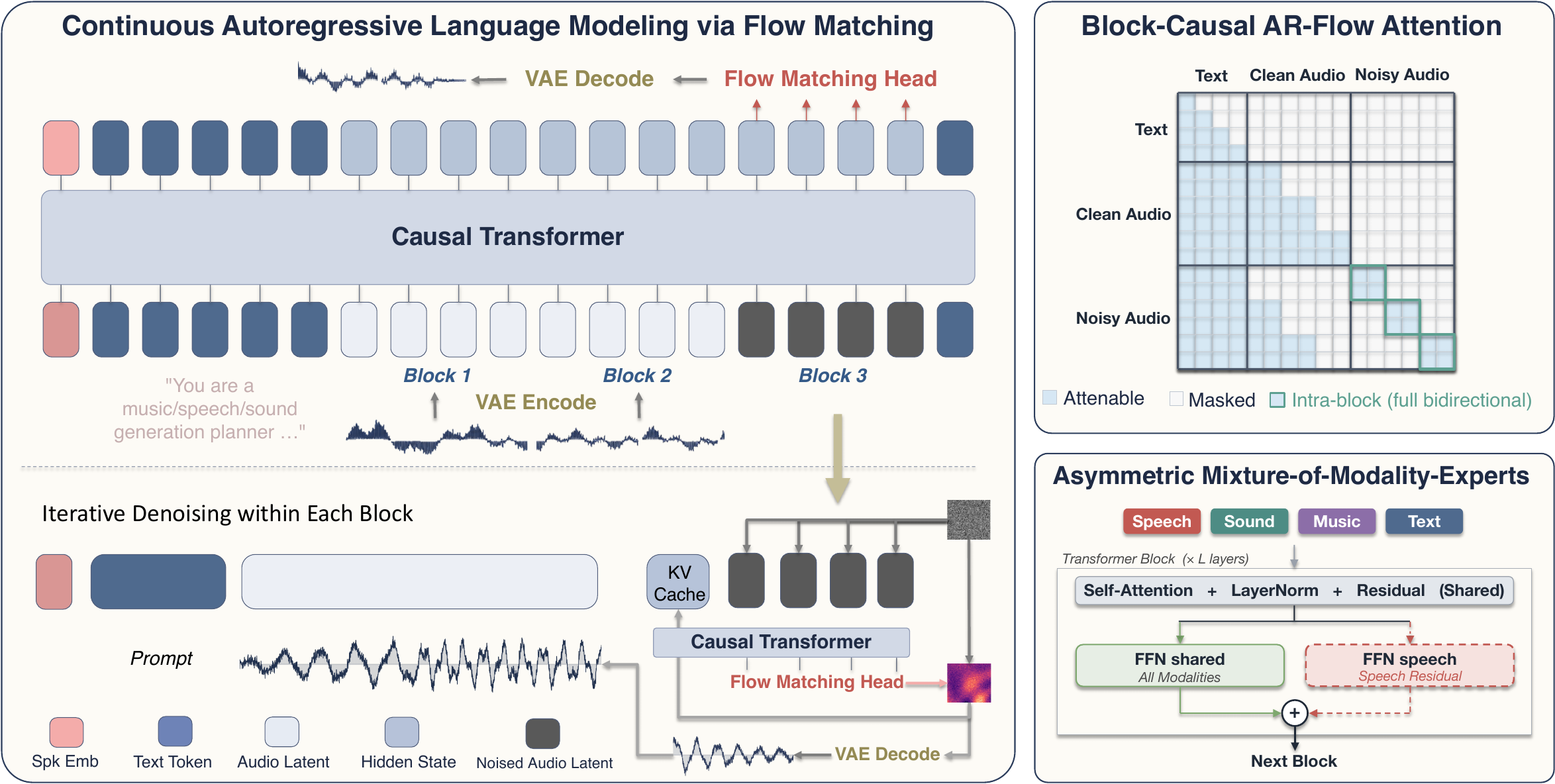}
    \caption{\textbf{Overview of AudioCALM.} \emph{Left:} a causal Transformer autoregresses over fixed-size blocks of continuous audio latents; a flow-matching head and a stop head are attached to its hidden states (top), and each block is produced by iterative denoising with KV-cache reuse (bottom). \emph{Top right:} the block-causal AR-Flow attention mask, causal across blocks and bidirectional within the active noisy block (red). \emph{Bottom right:} the Asymmetric Mixture-of-Modality-Experts layer, in which a deterministically routed speech-only FFN is added to the shared backbone.}
    \label{fig:overview}
    \vspace{-4mm}
\end{figure}
\vspace{-1mm}
\section{AudioCALM}
\label{sec:method}

\subsection{Overview}
\label{sec:overview}

Figure~\ref{fig:overview} illustrates the AudioCALM architecture. Section~\ref{sec:fmhead} introduces the flow-matching head that extends next-token prediction from a discrete softmax to a continuous rectified-flow velocity over VAE latents. Section~\ref{sec:arflow} develops the block-causal AR-Flow attention pattern that couples autoregressive commitment across blocks with bidirectional flow matching within a block, supporting streaming variable-length generation. Section~\ref{sec:asymmetry} addresses the speech--audio asymmetry through a description-style conditioning pipeline and the Asymmetric Mixture-of-Modality-Experts (A-MoME). Section~\ref{sec:training} specifies the training objective, exposure-bias regularizers, and autoregressive inference procedure.

\subsection{Continuous Autoregressive Language Modeling via Flow Matching}
\label{sec:fmhead}

\paragraph{Sequence formulation.}
A training example pairs a textual condition $c$ with continuous audio latents $x = (x_1, \ldots, x_L) \in \mathbb{R}^{L \times C}$ from a frozen VAE~\citep{vae, stableaudioopen}, with $L$ latent positions and $C$ channels per position, factorized autoregressively as $p(x \mid c) = \prod_i p(x_i \mid x_{<i}, c)$ where $x_{<i} = (x_1, \ldots, x_{i-1})$. For speech, $c$ is preceded by a single speaker embedding $s \in \mathbb{R}^{D_s}$ from a frozen speaker encoder (Spk Emb in Fig.~\ref{fig:overview}). Text is embedded via the LM's native token embedding; for each audio position $i$, a noisy latent $x_i^{(t)}$ at flow-matching timestep $t \in [0,1]$ enters the LM through a linear projection $\phi_{\text{in}} : \mathbb{R}^C \to \mathbb{R}^H$, summed with a learned position embedding $p_i$ and a sinusoidal timestep embedding $\tau(t)$,
\begin{equation}
\label{eq:audio-token}
e_i^{(t)} = \phi_{\text{in}}\!\big(x_i^{(t)}\big) + p_i + \tau(t),
\vspace{-1mm}
\end{equation}

where $t = 0$ marks a clean latent and $t \in (0,1)$ a noisy one. Text and audio embeddings are interleaved and processed by a shared transformer backbone $f_\theta$ initialized from a pretrained text LLM, yielding hidden states $h_i \in \mathbb{R}^H$.

\paragraph{Flow-matching head and training objective.}
At every audio position, a linear projection
\begin{equation}
\label{eq:fmhead}
v_\theta(h_i) := \phi_{\text{out}}(h_i) \in \mathbb{R}^C
\vspace{-1mm}
\end{equation}

maps the hidden state to a velocity in the same latent space as $x_i$, parameterizing each per-position conditional implicitly through the velocity field $v_\theta$. We zero-initialize $\phi_{\text{out}}$ so the velocity field starts at zero, preventing the untrained head from disturbing the backbone in early training. Following rectified flow~\citep{rectifiedflow} with a logit-normal timestep schedule~\citep{sd3}, we draw $t = \sigma(u)$ with $u \sim \mathcal{N}(0, 1)$ and $\sigma$ the logistic sigmoid, concentrating probability mass near $t \approx 0.5$ where denoising is hardest. With per-position noise $\epsilon_i \sim \mathcal{N}(0, I_C)$, the linear interpolation and target velocity are
\begin{equation}
\label{eq:linear-interpolation}
x_i^{(t)} = (1 - t)\,x_i + t\,\epsilon_i,
\qquad
v_i^{\star} := \epsilon_i - x_i,
\vspace{-1mm}
\end{equation}

and the training loss, averaged over the set $\mathcal{N}$ of noisy audio positions in the example, is
\begin{equation}
\label{eq:fm-loss}
\mathcal{L}_\text{FM}(\theta) = \mathbb{E}_{x,\,c,\,u,\,\epsilon}\!\left[\,\frac{1}{|\mathcal{N}|}\sum_{i \in \mathcal{N}} \big\|\, v_\theta(h_i) - v_i^{\star} \big\|_2^2\,\right].
\vspace{-1mm}
\end{equation}

Because $\phi_{\text{in}}$ embeds clean and noisy latents through the same projection, a clean prefix latent and a noisy latent at $t = 0$ produce identical embeddings, allowing a fully denoised block to be committed to the cache without train--inference mismatch.

\subsection{Block-Causal AR-Flow Attention}
\label{sec:arflow}

Combining autoregression with flow matching raises a structural question: causal attention serves AR commitment, while flow matching requires bidirectional access among the noisy latents being denoised together. AR-Flow resolves this by attending causally across blocks and fully within each block, yielding a single mask under which a flow-matching trajectory denoises one block while the LM commits the previous one.

\paragraph{Block-causal mask.}
Let $\tilde{x}_i^{(t)}$ denote a noisy latent at timestep $t$, and call a clean latent \emph{committed} once it has been denoised and written to the KV cache. At each generation step, the LM processes a packed sequence corresponding to one active noisy block (illustrated in Fig.~\ref{fig:overview} top-right):
\begin{equation}
\label{eq:packed-seq}
\big[\,c_1, \ldots, c_T,\; x_1, \ldots, x_i,\; \tilde{x}_{i+1}^{(t)}, \ldots, \tilde{x}_{i+B}^{(t)}\,\big],
\vspace{-1mm}
\end{equation}

formed by a $T$-token textual condition, $i$ committed clean latents, and a contiguous block of $B$ noisy latents sharing a block-level timestep $t$. The \emph{AR-Flow} mask enforces three rules:
\vspace{-1mm}
\begin{itemize}
\item every text token $c_j$ attends only to text tokens $c_{\le j}$;
\item every clean audio latent $x_k$ ($k \le i$) attends to all text tokens and to clean latents $x_{\le k}$;
\item every noisy latent $\tilde{x}_m^{(t)}$ ($i < m \le i+B$) attends to the entire textual condition, all committed clean latents $x_{1:i}$, and all noisy latents within its block, but not to any other noisy block.
\end{itemize}
\vspace{-1mm}
The mask is therefore causal across blocks and full within each block: the bidirectional intra-block context is what flow matching needs to denoise the block jointly, while the inter-block causality lets the cached prefix grow incrementally as blocks commit, so the model accommodates arbitrary length under the same mask at training and inference.

\paragraph{Single-pass teacher-forced training.}
A na\"ive realization would require one forward pass per noisy block, scaling linearly with the number of blocks per utterance and forfeiting the parallelism that makes LM training efficient. We instead collapse all blocks into a single pass over the extended packed sequence
\begin{equation}
\label{eq:extended-seq}
\big[\,c,\; x_1, \ldots, x_L \;\big|\; \tilde{x}_1^{(t_1)}, \ldots, \tilde{x}_L^{(t_L)}\,\big],
\vspace{-1mm}
\end{equation}

whose right half holds per-position noisy copies of the audio. Let $\beta(m)$ denote the index of the noisy block containing position $m$ and $b_s(m)$ its first position. The mask is generalized so that each $\tilde{x}_m^{(t_m)}$ attends to $c$, to the clean prefix $x_{1:\,b_s(m)-1}$ \emph{strictly preceding} its own block, and to all noisy positions in the same block---no noisy position ever attends to a clean counterpart inside its own block, so the training mask matches the inference mask of Eq.~\ref{eq:packed-seq} exactly. The two halves share position embeddings, so noisy copies occupy the same RoPE~\citep{roformer} positions as their clean counterparts and---by the $\phi_{\text{in}}$ equivalence in \S\ref{sec:fmhead}---train and inference interfaces coincide at $t = 0$. Timesteps are sampled once per block ($t_m = t_{m'}$ whenever $\beta(m) = \beta(m')$) and drawn independently across blocks, matching the inference schedule. The loss in Eq.~\ref{eq:fm-loss} is computed on the noisy half; the clean half exists only to populate the KV cache. The result: per-token training cost of a standard LM, no teacher-forcing/inference mismatch beyond drift (handled in \S\ref{sec:training}), and identical attention masks at training and generation.

\subsection{Modality-Asymmetric Specialization}
\label{sec:asymmetry}

The local--global text--audio mismatch identified in Section~\ref{sec:intro} manifests on two levels: a \emph{conditioning} gap, with speech tightly aligned to a transcript while sound and music are described by global captions, and a \emph{computation} gap, with speech demanding alignment-bound processing absent from the others. We close the former at the data level and the latter at the architecture level.

\paragraph{Description-style conditioning.}
Each clip is paired offline with two textual conditions. The \emph{short} variant is the corpus annotation---transcript for speech, original short caption for sound and music. The \emph{long-form} variant is produced by an audio-conditioned MLLM that consumes the clip together with the short variant and surfaces modality-relevant attributes the short variant cannot recover (speaker prosody and acoustic context for speech; acoustic events and scene structure for sound; instrumentation, tempo, and recurring motifs for music) as natural-language phrases grounded in the waveform; for speech, the verbatim transcript is wrapped in explicit content delimiters so local alignment is preserved. Both variants are sampled with equal probability per step, exposing the LM to terse and verbose prompts under one description-style interface usable with either at inference. The MLLM choice, prompt templates, decoding settings, and speech delimiter convention are in Appendix~\ref{sec:appendix-conditioning}.

\paragraph{Asymmetric Mixture-of-Modality-Experts.}
Inspired by modality-expert designs that share self-attention while specializing the feed-forward sub-layer per modality~\citep{vlmo, beit3, mot}, A-MoME keeps self-attention, layer normalization, residual connections, \emph{and} the main feed-forward network shared, introducing a single residual expert active only on speech positions. Let $h$ denote the hidden state at any token position and $m(h) \in \{\texttt{speech}, \texttt{sound}, \texttt{music}, \texttt{text}\}$ the modality tag; the modified feed-forward computation is
\begin{equation}
\label{eq:amome}
\mathrm{FFN}_{\text{A-MoME}}(h) =
\begin{cases}
\mathrm{FFN}_{\text{shared}}(h) + \mathrm{FFN}_{\text{speech}}(h), & m(h) = \texttt{speech},\\
\mathrm{FFN}_{\text{shared}}(h), & \text{otherwise},
\end{cases}
\end{equation}

where $\mathrm{FFN}_{\text{speech}}$ matches $\mathrm{FFN}_{\text{shared}}$ in form but has its own parameters, zero-initialized as with $\phi_{\text{out}}$ so A-MoME begins as a no-op and only departs from the shared backbone as training progresses. Modality tags are read from each position's source corpus, so the branch selection in Eq.~\ref{eq:amome} is deterministic with no gating network. Unlike symmetric modality-expert variants that allocate a separate FFN per modality, A-MoME adds only one extra FFN sub-layer per block, and non-speech inference is unaffected since sound, music, and text activate only $\mathrm{FFN}_{\text{shared}}$.

\subsection{Training and Inference}
\label{sec:training}

\paragraph{Training objective.}
The total objective combines the rectified-flow loss in Eq.~\ref{eq:fm-loss} on the noisy half of the extended sequence (Eq.~\ref{eq:extended-seq}) with a binary cross-entropy on the stop head (described below).

\paragraph{Closing the train--inference gap.}
In inference, each committed block is the denoised latent $\hat{x}$ from the previous step's flow-matching trajectory, re-embedded as a clean prefix for the next step. Drift from the true clean latent accumulates across blocks, opening a gap that pure teacher forcing does not capture. We close it with two complementary regularizers:
\begin{itemize}
\item \textbf{Per-block clean-prefix noise} ($\sigma_\text{clean}$). We perturb only the input embeddings $\phi_{\text{in}}(x_i)$ of clean prefix tokens with Gaussian noise; the noise scale ramps linearly from $0$ at the first block to $\sigma_\text{clean}$ at the last block of the utterance and is held constant within each block. The interpolation $x_i^{(t)}$ and target $v_i^{\star}$ remain computed from the unperturbed $x_i$, so the supervision is unchanged and the perturbation only simulates per-block drift in the cached prefix.
\item \textbf{Exposure-bias perturbation} ($\gamma_\text{exp}$). For each sample we draw $\alpha \sim \mathcal{U}[0, \gamma_\text{exp}]$ and form a perturbed clean target $\tilde{x}_i = x_i + \alpha\,\xi_i$ with $\xi_i \sim \mathcal{N}(0, I_C)$. Both the prefix embedding and the target $v_i^{\star} = \epsilon_i - \tilde{x}_i$ are computed from $\tilde{x}_i$, training the model to recover from a slightly noisy commitment.
\end{itemize}

\paragraph{Autoregressive inference.}
AR-Flow accommodates arbitrary lengths by construction, so AudioCALM needs no external duration predictor. A lightweight binary stop head $g_\text{stop} : \mathbb{R}^H \to (0, 1)$ on the clean hidden states is trained with binary cross-entropy whose targets ramp linearly from $0$ to $1$ over the last $K_\text{stop}$ tokens of each clip, so the model learns to anticipate clip end. At inference, each block is denoised by Euler integration of the learned velocity field~\citep{rectifiedflow} and committed to the KV cache; $g_\text{stop}$ is then queried at every position of the new block, and the first position exceeding a fixed threshold $\tau_{\text{stop}}$ cuts the output, yielding token- rather than block-level termination granularity.

\vspace{-1mm}
\section{Experiments}
\label{sec:exp}

\subsection{Experimental Setup}
\label{subsec:setup}

\paragraph{Training Data.}
We jointly train AudioCALM on three corpora covering the universal audio setting. For speech, we combine \textsc{LibriTTS}~\citep{libritts} with the English subset of \textsc{Emilia}~\citep{emilia}; for general sound, we aggregate the audio of \textsc{VGGSound}~\citep{vggsound}, \textsc{AudioCaps}~\citep{audiocaps}, and \textsc{WavCaps}~\citep{wavcaps}; for music, we combine \textsc{FMA}~\citep{fma} with \textsc{MTG-Jamendo}~\citep{mtgjamendo}. Only the official train splits are used; LibriTTS \emph{test-clean}, the AudioCaps and Song-Describer evaluation sets, and SeedTTS-eval are held out to avoid test-set contamination. All audio is resampled to $44.1$\,kHz, and every clip is re-captioned with \textsc{Gemini~3~Pro}\footnote{\url{https://deepmind.google/technologies/gemini/}} into long-form natural-language descriptions through the description-style pipeline of Section~\ref{sec:asymmetry}, with the original transcript or tag retained as a parallel short-form conditioning channel. Corpus details are in Appendix~\ref{sec:appendix-data}, and the verbatim \textsc{Gemini~3~Pro} prompts in Appendix~\ref{sec:appendix-conditioning}.

\paragraph{Evaluation Benchmarks.}
We follow standard evaluation protocols for each domain. Text-to-speech is evaluated on the full \textsc{LibriTTS} \emph{test-clean} split and the \textit{en} split of \textsc{SeedTTS-eval}~\citep{seedtts}; text-to-sound on \textsc{AudioCaps} and text-to-music on the held-out \textsc{Song-Describer}~\citep{songdescriber} captions.

\paragraph{Evaluation Metrics.}
We adopt both objective and subjective metrics. For speech, \textbf{WER} (Whisper-large-v3~\citep{whisper}) measures intelligibility, and \textbf{SIM} (cosine similarity between \textsc{WavLM-Large}~\citep{wavlm} speaker-verification embeddings of the generation and the reference) measures speaker fidelity. For sound and music, \textbf{FAD}~\citep{fad}---the Fr\'echet distance in VGGish~\citep{vggish} embedding space---measures distributional fidelity, and LAION-\textbf{CLAP}~\citep{clap} text--audio cosine similarity measures prompt adherence. Subjective evaluation uses \textbf{MOS} (overall quality) for speech and \textbf{MOS-Q}/\textbf{MOS-T} (quality, text relevance) for sound and music, on $N{=}50$ samples per system; full definitions and protocol are in Appendix~\ref{sec:appendix-metrics}.

\paragraph{Baselines.}
We compare against modality-specific and unified baselines spanning the dominant generative paradigms in each domain. \emph{Modality-specific:} \textsc{F5-TTS}~\citep{f5tts} and \textsc{CosyVoice\,3.0}~\citep{cosyvoice3} for speech; \textsc{AudioLDM\,2-Large}~\citep{audioldm2}, \textsc{TangoFlux}~\citep{tangoflux}, and \textsc{Stable Audio Open}~\citep{stableaudioopen} for sound; \textsc{MusicGen-Large}~\citep{musicgen} and \textsc{Stable Audio Open} for music. CosyVoice\,3.0, TangoFlux, and Stable Audio Open are the current open SOTA in their respective domains. \emph{Unified:} \textsc{UniAudio}~\citep{uniaudio}, \textsc{UniMoE-Audio}~\citep{unimoeaudio}, \textsc{UniFlow-Audio}~\citep{uniflowaudio}, and \textsc{Ming-omni}~\citep{mingomni}, one representative per architectural line of Section~\ref{sec:related}. All baselines are evaluated from their official checkpoints under the recommended decoding configurations.

\paragraph{Implementation Details.}
The backbone is initialized from \textsc{Qwen3-1.7B}~\citep{qwen3}; audio is encoded by our own continuous VAE on $44.1$\,kHz waveforms. We train for $300$\,k steps on A800${\times}8$ GPUs under FSDP HYBRID-SHARD with AdamW ($\beta_1{=}0.9$, $\beta_2{=}0.95$, weight decay $0$) at a constant learning rate of $1{\times}10^{-4}$ following $2{,}000$ warmup steps, with a global batch of $\sim\!64$\,k tokens (8\,k tokens per GPU) and gradient clipping at $1.0$, optimizing the rectified-flow objective. We apply $10\%$ caption dropout for classifier-free guidance with scale $w{=}3$. Inference uses $K_\text{flow}{=}24$ rectified-flow steps per block and an AR-Flow block size of $B{=}1.0$\,s, which corresponds to $11$ VAE latent tokens at the bottleneck's $\approx\!10.75$\,Hz rate (Appendix~\ref{sec:appendix-impl}). VAE architecture and training are in Appendix~\ref{sec:appendix-impl}, and the full hyperparameter ledger is in Appendix~\ref{sec:appendix-hparams}.

\subsection{Universal Audio Generation Results}
\label{subsec:main-results}

Tables~\ref{tab:tts} and~\ref{tab:t2a-music} compare AudioCALM against the two baseline families on the three target domains. Across all three domains, AudioCALM---a single set of weights---achieves results comparable to or better than the best modality-specific system, while clearly outperforming the unified baselines. AudioCALM also streams natively without an external duration predictor, a property the non-AR flow-matching baselines lack.

\begin{table}[t]
    \vspace{-2mm}
\centering
\small
\setlength{\tabcolsep}{4pt}
\caption{Text-to-speech evaluation on \textsc{LibriTTS} \emph{test-clean} and the \emph{en} split of \textsc{SeedTTS-eval}. MOS is reported as mean$\pm$std. Best per column in \textbf{bold}, second-best \underline{underlined}.}
\label{tab:tts}
\begin{tabular*}{\linewidth}{@{\extracolsep{\fill}}lcccccc}
\toprule
& \multicolumn{3}{c}{LibriTTS} & \multicolumn{3}{c}{SeedTTS-eval (en)} \\
\cmidrule(lr){2-4}\cmidrule(lr){5-7}
Model & WER$\downarrow$ & SIM$\uparrow$ & MOS$\uparrow$ & WER$\downarrow$ & SIM$\uparrow$ & MOS$\uparrow$ \\
\midrule
\multicolumn{7}{l}{\emph{Modality-specific baselines}} \\
F5-TTS         & 0.033 & 0.616 & $3.85_{\pm 0.08}$ & 0.018 & 0.648 & $3.78_{\pm 0.09}$ \\
CosyVoice\,3.0 & \underline{0.022} & \textbf{0.697} & $\underline{3.96_{\pm 0.07}}$ & 0.015 & \textbf{0.695} & $\underline{3.88_{\pm 0.08}}$ \\
\midrule
\multicolumn{7}{l}{\emph{Unified baselines}} \\
UniAudio       & 0.120 & 0.265 & $3.30_{\pm 0.11}$ & 0.113 & 0.363 & $3.22_{\pm 0.12}$ \\
UniMoE-Audio   & 0.078 & 0.361 & $3.52_{\pm 0.09}$ & 0.019 & 0.573 & $3.72_{\pm 0.08}$ \\
UniFlow-Audio  & 0.032 & 0.570 & $3.50_{\pm 0.10}$ & 0.058 & 0.573 & $3.45_{\pm 0.10}$ \\
Ming-omni-TTS  & 0.025 & 0.553 & $3.82_{\pm 0.08}$ & \underline{0.013} & 0.633 & $3.80_{\pm 0.07}$ \\
\midrule
\textbf{AudioCALM} (ours) & \textbf{0.020} & \underline{0.668} & $\mathbf{4.02_{\pm 0.06}}$ & \textbf{0.011} & \underline{0.672} & $\mathbf{3.95_{\pm 0.07}}$ \\
\bottomrule
\end{tabular*}
\end{table}

\begin{table}[t]
\centering
\small
\setlength{\tabcolsep}{4pt}
\caption{Text-to-sound (AudioCaps) and text-to-music (Song-Describer) evaluation. ``--'' marks domains a modality-specific model is not trained for; UniMoE-Audio additionally does not support text-to-sound.}
\label{tab:t2a-music}
\resizebox{\linewidth}{!}{%
\begin{tabular}{lcccccccc}
\toprule
& \multicolumn{4}{c}{Text-to-sound (AudioCaps)} & \multicolumn{4}{c}{Text-to-music (Song-Describer)} \\
\cmidrule(lr){2-5}\cmidrule(lr){6-9}
Model & FAD$\downarrow$ & CLAP$\uparrow$ & MOS-Q$\uparrow$ & MOS-T$\uparrow$ & FAD$\downarrow$ & CLAP$\uparrow$ & MOS-Q$\uparrow$ & MOS-T$\uparrow$ \\
\midrule
\multicolumn{9}{l}{\emph{Modality-specific baselines}} \\
AudioLDM\,2-Large & 5.36 & 0.22 & $3.25_{\pm 0.10}$ & $3.10_{\pm 0.11}$ & --    & --   & --    & --    \\
TangoFlux         & 2.70  & 0.36 & $3.82_{\pm 0.07}$ & $3.85_{\pm 0.08}$ & --    & --   & --    & --    \\
Stable Audio Open & 4.13  & 0.25 & $3.65_{\pm 0.08}$ & $3.45_{\pm 0.09}$ & 2.23 & 0.32 & $3.95_{\pm 0.07}$ & $3.85_{\pm 0.08}$ \\
MusicGen-Large    & --    & --   & --    & --    & 5.28  & 0.19 & $3.65_{\pm 0.08}$ & $3.45_{\pm 0.10}$ \\
\midrule
\multicolumn{9}{l}{\emph{Unified baselines}} \\
UniAudio          & 6.64  & 0.13 & $3.20_{\pm 0.11}$ & $2.95_{\pm 0.13}$ & 11.25 & 0.06 & $2.80_{\pm 0.14}$ & $2.65_{\pm 0.15}$ \\
UniMoE-Audio      & --    & --   & --    & --    & 3.71  & 0.22 & $3.80_{\pm 0.08}$ & $3.60_{\pm 0.09}$ \\
UniFlow-Audio     & 4.22  & 0.35 & $3.62_{\pm 0.08}$ & $3.80_{\pm 0.08}$ & 6.39  & 0.15 & $3.45_{\pm 0.10}$ & $3.25_{\pm 0.11}$ \\
Ming-omni         & 2.46  & 0.27 & $3.85_{\pm 0.07}$ & $3.60_{\pm 0.09}$ & 7.98  & 0.07 & $3.25_{\pm 0.11}$ & $2.92_{\pm 0.13}$ \\
\midrule
\textbf{AudioCALM} (ours) & \textbf{1.95} & \textbf{0.37} & $\mathbf{3.98_{\pm 0.06}}$ & $\mathbf{3.95_{\pm 0.07}}$ & \textbf{2.02} & \textbf{0.36} & $\mathbf{3.99_{\pm 0.06}}$ & $\mathbf{3.92_{\pm 0.07}}$ \\
\bottomrule
\end{tabular}%
}
\vspace{-4mm}
\end{table}

\paragraph{Zero-Shot TTS.}
Both benchmarks follow the standard zero-shot TTS protocol---a speaker embedding extracted from a $3$\,s reference utterance is prepended as the first input token---and report WER (intelligibility), SIM (speaker fidelity), and MOS (overall quality). AudioCALM achieves the lowest WER on both \textsc{LibriTTS} \emph{test-clean} ($0.020$) and \textsc{SeedTTS-eval}~\emph{en} ($0.011$) and the highest MOS on both ($4.02$ and $3.95$), surpassing every modality-specific and unified baseline. On speaker similarity, it trails only CosyVoice\,3.0 (LibriTTS: $0.668$ vs.\ $0.697$; SeedTTS-eval: $0.672$ vs.\ $0.695$), the only columns where a specialist outperforms us---a gap consistent with CosyVoice\,3.0's much larger English-only speech corpus and dedicated speaker-fidelity objective, neither of which a unified, multi-modal system would be expected to match.

\paragraph{Sound and Music Generation.}
On \textsc{AudioCaps}, AudioCALM dominates every column, lowering FAD from the strongest sound generation specialist (TangoFlux, $2.70$) to $1.95$---a $28\%$ relative reduction---and edging LAION-CLAP from $0.36$ to $0.37$. On \textsc{Song-Describer}, AudioCALM likewise dominates: lowest FAD ($2.02$ vs.\ Stable Audio Open's $2.23$) and highest text--audio CLAP ($0.36$ vs.\ $0.32$). Both subjective axes (MOS-Q and MOS-T) also place AudioCALM first on every non-speech track.

\paragraph{Cross-Modal Interference.}
Existing universal audio generation systems exhibit clear \emph{modality collapse}: \textsc{UniAudio} trails the SOTA by a large margin on every modality, \textsc{UniFlow-Audio} loses $\geq 2.8\times$ FAD on music ($6.39$ vs.\ $2.23$), and \textsc{Ming-omni} breaks down on music (FAD $7.98$, CLAP $0.07$). AudioCALM is the only system in Tables~\ref{tab:tts} and~\ref{tab:t2a-music} that ranks first or second on \emph{every} column we report, supporting the claim that continuous-latent autoregression combined with description-style conditioning eliminates the cross-modal interference that has defined unified audio generation to date.

\subsection{Ablation Studies}
\label{subsec:ablation}

We ablate AudioCALM's three core designs on text-to-speech (\textsc{LibriTTS}~\emph{test-clean}; Sp.\ WER, SIM), text-to-sound (\textsc{AudioCaps}; Sn.\ FAD, CLAP), and text-to-music (\textsc{Song-Describer}; Mu.\ FAD, CLAP). The cumulative variants (a)--(e) in Table~\ref{tab:ablation} share the joint training of Section~\ref{subsec:setup}, so each column reflects per-modality fidelity \emph{and} any cross-modal crowd-out; the two single-modality rows at the top use variant (b)'s architecture on one modality, isolating crowd-out from architectural effects. Backbone, CFG, block-size, and inference-step sweeps are deferred to Appendices~\ref{sec:appendix-backbone-scale}--\ref{sec:appendix-step-sweep}.

\begin{table}[t]
\centering
\vspace{-2mm}
\small
\caption{Ablation on text-to-speech (\textsc{LibriTTS}~\emph{test-clean}), text-to-sound (\textsc{AudioCaps}), and text-to-music (\textsc{Song-Describer}). Top: single-modality specialists with variant (b)'s architecture, isolating crowd-out from architectural effects. Bottom: cumulative joint-training variants (a)--(e) sharing the Qwen3-1.7B backbone of Section~\ref{subsec:setup}; (a) shares the same VAE encoder/decoder as (b)--(e) but applies vector quantization to a discrete codebook at the bottleneck in place of the continuous flow-matching head, so the (a)$\to$(b) comparison isolates the head choice from the codec.}
\label{tab:ablation}
\resizebox{\linewidth}{!}{%
\begin{tabular}{lcccccc}
\toprule
& \multicolumn{2}{c}{Speech (LibriTTS)} & \multicolumn{2}{c}{Sound (AudioCaps)} & \multicolumn{2}{c}{Music (Song-Describer)} \\
\cmidrule(lr){2-3}\cmidrule(lr){4-5}\cmidrule(lr){6-7}
Variant & WER$\downarrow$ & SIM$\uparrow$ & FAD$\downarrow$ & CLAP$\uparrow$ & FAD$\downarrow$ & CLAP$\uparrow$ \\
\midrule
\multicolumn{7}{l}{\emph{Single-modality specialists (variant (b) architecture, no joint training)}} \\
Speech-only training         & 0.022 & 0.628 & --   & --   & --   & --   \\
Non-speech-only training     & --   & --   & 2.45  & 0.34  & 2.55  & 0.30  \\
\midrule
\multicolumn{7}{l}{\emph{Joint speech/sound/music training, cumulative ablation}} \\
(a) Discrete-token AR baseline (no flow head)   & 0.040 & 0.560 & 4.50 & 0.22 & 4.80 & 0.20 \\
(b) Continuous AR + flow head, raw transcripts  & 0.024 & 0.620 & 3.30 & 0.30 & 3.45 & 0.26 \\
(c) (b) + description-style reframing           & 0.023 & 0.625 & 2.70 & 0.33 & 2.85 & 0.29 \\
(d) (c) + symmetric MoME (3 modality experts)   & 0.022 & 0.660 & 2.30 & 0.35 & 2.40 & 0.33 \\
(e) (c) + \textbf{A-MoME (speech residual only)} & \textbf{0.020} & \textbf{0.668} & \textbf{1.95} & \textbf{0.37} & \textbf{2.02} & \textbf{0.36} \\
\bottomrule
\end{tabular}%
}
\vspace{-4mm}
\end{table}

\paragraph{Asymmetric Crowd-Out.}
The two specialist rows vs.\ joint variant (b)---same architecture, only the training mixture differs---directly verify the asymmetric mismatch claimed in Section~\ref{sec:intro}. Adding non-speech data to a speech-only run leaves speech metrics within noise (Sp.\ WER $0.022\!\to\!0.024$; SIM $0.628\!\to\!0.620$), whereas adding speech to a non-speech-only run inflates Sn.\ FAD $2.45\!\to\!3.30$ and Mu.\ FAD $2.55\!\to\!3.45$ ($+35\%$ each) and drops both CLAP scores by $10$--$13\%$. The crowd-out is thus directional---non-speech pays a $\sim\!4\times$ larger fidelity tax---motivating the asymmetric data and architecture fixes below over uniform capacity across the three modalities.

\paragraph{Continuous Flow-Matching Head.}
Replacing (a)'s discrete codec head with (b)'s continuous flow-matching head is the single largest contributor across all three domains (Sp.\ WER $0.040{\to}0.024$, SIM $0.560{\to}0.620$; Sn.\ FAD $4.50{\to}3.30$, CLAP $0.22{\to}0.30$; Mu.\ FAD $4.80{\to}3.45$, CLAP $0.20{\to}0.26$). The codec bottleneck is therefore the dominant fidelity cap on all three sides, and continuous-latent autoregression alone closes most of the gap to non-AR flow-matching baselines.

\paragraph{Description-Style Reframing.}
Rewriting raw transcripts and captions into long-form descriptions (c vs.\ b) lifts Sn.\ FAD/CLAP $3.30$/$0.30\!\to\!2.70$/$0.33$ and Mu.\ FAD/CLAP $3.45$/$0.26\!\to\!2.85$/$0.29$ while leaving speech metrics within noise (Sp.\ WER $0.024\!\to\!0.023$, SIM $0.620\!\to\!0.625$). The non-speech gain stems from harmonising the conditioning surface across modalities rather than sacrificing speech accuracy: once transcripts and captions look alike to the LM, the shared self-attention layers escape the local-vs-global tug-of-war of raw-transcript joint training.

\paragraph{A-MoME vs.\ Symmetric MoME.}
Symmetric three-expert MoME (d) reaches Sp.\ $0.022$/$0.660$, Sn.\ $2.30$/$0.35$, Mu.\ $2.40$/$0.33$, but duplicates the FFN width across \emph{every} block. A-MoME (e), adding one speech-only residual FFN, improves all six metrics to $\mathbf{0.020}$/$\mathbf{0.668}$/$\mathbf{1.95}$/$\mathbf{0.37}$/$\mathbf{2.02}$/$\mathbf{0.36}$ with markedly fewer extra parameters and \emph{zero} inference overhead on non-speech inputs---speech gets a dedicated branch while sound and music stop competing for shared FFN capacity. Capacity allocation that tracks the asymmetric source of the mismatch, therefore beats one that tracks the modality count.



\vspace{-1mm}
\section{Conclusion}
\label{sec:con}

We presented \textbf{AudioCALM}, a universal audio generation framework instantiating \emph{Continuous Autoregressive Language Modeling (CALM)}: a single autoregressive LM parameterizes per-position rectified-flow velocities via a thin flow-matching head, paired with a block-causal AR-Flow attention pattern for streaming, variable-length generation of speech, sound, and music. We further identified an asymmetric mismatch between locally aligned text-to-speech and globally conditioned text-to-audio, resolving it with a unified description-style conditioning interface and A-MoME, whose granularity tracks this mismatch rather than the modality count. As a single set of weights, AudioCALM matches state-of-the-art dedicated systems on all three benchmarks. Open questions remain around corpus coverage, backbone scale, and long-form generation; deployment additionally requires provenance and consent safeguards against voice-cloning and synthetic-media risks (see Appendices~\ref{sec:appendix-limitations},~\ref{sec:appendix-broader}, and~\ref{sec:appendix-safeguards}).


{
\small
\bibliographystyle{plainnat}
\bibliography{neurips_2026}

@article{uniaudio,
  title={Uniaudio: An audio foundation model toward universal audio generation},
  author={Yang, Dongchao and Tian, Jinchuan and Tan, Xu and Huang, Rongjie and Liu, Songxiang and Chang, Xuankai and Shi, Jiatong and Zhao, Sheng and Bian, Jiang and Wu, Xixin and others},
  journal={arXiv preprint arXiv:2310.00704},
  year={2023}
}

@article{uniaudio15,
  title={Uniaudio 1.5: Large language model-driven audio codec is a few-shot audio task learner},
  author={Yang, Dongchao and Guo, Haohan and Wang, Yuanyuan and Huang, Rongjie and Li, Xiang and Tan, Xu and Wu, Xixin and Meng, Helen},
  journal={Advances in Neural Information Processing Systems},
  volume={37},
  pages={56802--56827},
  year={2024}
}

@article{cosyvoice2,
    title={Cosyvoice 2: Scalable streaming speech synthesis with large language models},
  author={Du, Zhihao and Wang, Yuxuan and Chen, Qian and Shi, Xian and Lv, Xiang and Zhao, Tianyu and Gao, Zhifu and Yang, Yexin and Gao, Changfeng and Wang, Hui and others},
  journal={arXiv preprint arXiv:2412.10117},
  year={2024}
}

@article{maskgct,
  title={Maskgct: Zero-shot text-to-speech with masked generative codec transformer},
  author={Wang, Yuancheng and Zhan, Haoyue and Liu, Liwei and Zeng, Ruihong and Guo, Haotian and Zheng, Jiachen and Zhang, Qiang and Zhang, Xueyao and Zhang, Shunsi and Wu, Zhizheng},
  journal={arXiv preprint arXiv:2409.00750},
  year={2024}
}

@article{audiobox,
    title={Audiobox: Unified audio generation with natural language prompts},
  author={Vyas, Apoorv and Shi, Bowen and Le, Matthew and Tjandra, Andros and Wu, Yi-Chiao and Guo, Baishan and Zhang, Jiemin and Zhang, Xinyue and Adkins, Robert and Ngan, William and others},
  journal={arXiv preprint arXiv:2312.15821},
  year={2023}
}

@inproceedings{stableaudioopen,
  title={Stable audio open},
  author={Evans, Zach and Parker, Julian D and Carr, CJ and Zukowski, Zack and Taylor, Josiah and Pons, Jordi},
  booktitle={ICASSP 2025-2025 IEEE International Conference on Acoustics, Speech and Signal Processing (ICASSP)},
  pages={1--5},
  year={2025},
  organization={IEEE}
}

@inproceedings{speechgpt,
    title={Speechgpt: Empowering large language models with intrinsic cross-modal conversational abilities},
    author={Zhang, Dong and Li, Shimin and Zhang, Xin and Zhan, Jun and Wang, Pengyu and Zhou, Yaqian and Qiu, Xipeng},
    booktitle={Findings of the Association for Computational Linguistics: EMNLP 2023},
    pages={15757--15773},
    year={2023}
}

@article{audiopalm,
  title={Audiopalm: A large language model that can speak and listen},
  author={Rubenstein, Paul K and Asawaroengchai, Chulayuth and Nguyen, Duc Dung and Bapna, Ankur and Borsos, Zal{\'a}n and Quitry, F{\'e}lix de Chaumont and Chen, Peter and Badawy, Dalia El and Han, Wei and Kharitonov, Eugene and others},
  journal={arXiv preprint arXiv:2306.12925},
  year={2023}
}

@article{moshi,
  title={Moshi: a speech-text foundation model for real-time dialogue},
  author={D{\'e}fossez, Alexandre and Mazar{\'e}, Laurent and Orsini, Manu and Royer, Am{\'e}lie and P{\'e}rez, Patrick and J{\'e}gou, Herv{\'e} and Grave, Edouard and Zeghidour, Neil},
  journal={arXiv preprint arXiv:2410.00037},
  year={2024}
}

@article{audiolm,
  title={Audiolm: a language modeling approach to audio generation},
  author={Borsos, Zal{\'a}n and Marinier, Rapha{\"e}l and Vincent, Damien and Kharitonov, Eugene and Pietquin, Olivier and Sharifi, Matt and Roblek, Dominik and Teboul, Olivier and Grangier, David and Tagliasacchi, Marco and others},
  journal={IEEE/ACM transactions on audio, speech, and language processing},
  volume={31},
  pages={2523--2533},
  year={2023},
  publisher={IEEE}
}

@article{valle,
    title={Neural codec language models are zero-shot text to speech synthesizers},
  author={Wang, Chengyi and Chen, Sanyuan and Wu, Yu and Zhang, Ziqiang and Zhou, Long and Liu, Shujie and Chen, Zhuo and Liu, Yanqing and Wang, Huaming and Li, Jinyu and others},
  journal={arXiv preprint arXiv:2301.02111},
  year={2023}
}

@article{musicgen,
  title={Simple and controllable music generation},
  author={Copet, Jade and Kreuk, Felix and Gat, Itai and Remez, Tal and Kant, David and Synnaeve, Gabriel and Adi, Yossi and D{\'e}fossez, Alexandre},
  journal={Advances in neural information processing systems},
  volume={36},
  pages={47704--47720},
  year={2023}
}

@article{audiogen,
  title={Audiogen: Textually guided audio generation},
  author={Kreuk, Felix and Synnaeve, Gabriel and Polyak, Adam and Singer, Uriel and D{\'e}fossez, Alexandre and Copet, Jade and Parikh, Devi and Taigman, Yaniv and Adi, Yossi},
  journal={arXiv preprint arXiv:2209.15352},
  year={2022}
}

@article{encodec,
  title={High fidelity neural audio compression},
  author={D{\'e}fossez, Alexandre and Copet, Jade and Synnaeve, Gabriel and Adi, Yossi},
  journal={arXiv preprint arXiv:2210.13438},
  year={2022}
}

@article{dac,
    title={High-fidelity audio compression with improved rvqgan},
  author={Kumar, Rithesh and Seetharaman, Prem and Luebs, Alejandro and Kumar, Ishaan and Kumar, Kundan},
  journal={Advances in Neural Information Processing Systems},
  volume={36},
  pages={27980--27993},
  year={2023}
}

@article{audioldm2,
    title={Audioldm 2: Learning holistic audio generation with self-supervised pretraining},
  author={Liu, Haohe and Yuan, Yi and Liu, Xubo and Mei, Xinhao and Kong, Qiuqiang and Tian, Qiao and Wang, Yuping and Wang, Wenwu and Wang, Yuxuan and Plumbley, Mark D},
  journal={IEEE/ACM Transactions on Audio, Speech, and Language Processing},
  volume={32},
  pages={2871--2883},
  year={2024},
  publisher={IEEE}
}

@article{audiolcm,
  title={Audiolcm: Text-to-audio generation with latent consistency models},
  author={Liu, Huadai and Huang, Rongjie and Liu, Yang and Cao, Hengyuan and Wang, Jialei and Cheng, Xize and Zheng, Siqi and Zhao, Zhou},
  journal={arXiv preprint arXiv:2406.00356},
  year={2024}
}

@article{flashaudio,
  title={Flashaudio: Rectified flows for fast and high-fidelity text-to-audio generation},
  author={Liu, Huadai and Wang, Jialei and Huang, Rongjie and Liu, Yang and Lu, Heng and Zhao, Zhou and Xue, Wei},
  journal={arXiv preprint arXiv:2410.12266},
  year={2024}
}

@article{vitttts,
  title={Vit-tts: Visual text-to-speech with scalable diffusion transformer},
  author={Liu, Huadai and Huang, Rongjie and Lin, Xuan and Xu, Wenqiang and Zheng, Maozong and Chen, Hong and He, Jinzheng and Zhao, Zhou},
  journal={arXiv preprint arXiv:2305.12708},
  year={2023}
}

@article{thinksound,
  title={Thinksound: Chain-of-thought reasoning in multimodal large language models for audio generation and editing},
  author={Liu, Huadai and Luo, Kaicheng and Wang, Jialei and Wang, Wen and Chen, Qian and Zhao, Zhou and Xue, Wei},
  journal={arXiv preprint arXiv:2506.21448},
  year={2025}
}

@article{omniaudio,
  title={OmniAudio: Generating spatial audio from 360-degree video},
  author={Liu, Huadai and Luo, Tianyi and Luo, Kaicheng and Jiang, Qikai and Sun, Peiwen and Wang, Jialei and Huang, Rongjie and Chen, Qian and Wang, Wen and Li, Xiangtai},
  journal={arXiv preprint arXiv:2504.14906},
  year={2025}
}

@article{prismaudio,
  title={PrismAudio: Decomposed chain-of-thoughts and multi-dimensional rewards for video-to-audio generation},
  author={Liu, Huadai and Luo, Kaicheng and Wang, Wen and Chen, Qian and Sun, Peiwen and Huang, Rongjie and Li, Xiangang and Ye, Jieping and Xue, Wei},
  journal={arXiv preprint arXiv:2511.18833},
  year={2025}
}

@article{voicebox,
    title={Voicebox: Text-guided multilingual universal speech generation at scale},
  author={Le, Matthew and Vyas, Apoorv and Shi, Bowen and Karrer, Brian and Sari, Leda and Moritz, Rashel and Williamson, Mary and Manohar, Vimal and Adi, Yossi and Mahadeokar, Jay and others},
  journal={Advances in neural information processing systems},
  volume={36},
  pages={14005--14034},
  year={2023}
}

@inproceedings{matchatts,
  title={Matcha-TTS: A fast TTS architecture with conditional flow matching},
  author={Mehta, Shivam and Tu, Ruibo and Beskow, Jonas and Sz{\'e}kely, {\'E}va and Henter, Gustav Eje},
  booktitle={ICASSP 2024-2024 IEEE International Conference on Acoustics, Speech and Signal Processing (ICASSP)},
  pages={11341--11345},
  year={2024},
  organization={IEEE}
}

@article{naturalspeech3,
    title={Naturalspeech 3: Zero-shot speech synthesis with factorized codec and diffusion models},
  author={Ju, Zeqian and Wang, Yuancheng and Shen, Kai and Tan, Xu and Xin, Detai and Yang, Dongchao and Liu, Yanqing and Leng, Yichong and Song, Kaitao and Tang, Siliang and others},
  journal={arXiv preprint arXiv:2403.03100},
  year={2024}
}

@article{rectifiedflow,
    title={Flow straight and fast: Learning to generate and transfer data with rectified flow},
  author={Liu, Xingchao and Gong, Chengyue and Liu, Qiang},
  journal={arXiv preprint arXiv:2209.03003},
  year={2022}
}

@inproceedings{sd3,
    title={Scaling rectified flow transformers for high-resolution image synthesis},
  author={Esser, Patrick and Kulal, Sumith and Blattmann, Andreas and Entezari, Rahim and M{\"u}ller, Jonas and Saini, Harry and Levi, Yam and Lorenz, Dominik and Sauer, Axel and Boesel, Frederic and others},
  booktitle={Forty-first international conference on machine learning},
  year={2024}
}

@article{cfg,
    title={Classifier-free diffusion guidance},
  author={Ho, Jonathan and Salimans, Tim},
  journal={arXiv preprint arXiv:2207.12598},
  year={2022}
}

@inproceedings{givt,
    title={Givt: Generative infinite-vocabulary transformers},
  author={Tschannen, Michael and Eastwood, Cian and Mentzer, Fabian},
  booktitle={European Conference on Computer Vision},
  pages={292--309},
  year={2024},
  organization={Springer}
}

@article{mar,
    title={Autoregressive image generation without vector quantization},
  author={Li, Tianhong and Tian, Yonglong and Li, He and Deng, Mingyang and He, Kaiming},
  journal={Advances in Neural Information Processing Systems},
  volume={37},
  pages={56424--56445},
  year={2024}
}

@article{ditar,
    title={Ditar: Diffusion transformer autoregressive modeling for speech generation},
  author={Jia, Dongya and Chen, Zhuo and Chen, Jiawei and Du, Chenpeng and Wu, Jian and Cong, Jian and Zhuang, Xiaobin and Li, Chumin and Wei, Zhen and Wang, Yuping and others},
  journal={arXiv preprint arXiv:2502.03930},
  year={2025}
}

@article{uniflowaudio,
    title={Uniflow-audio: Unified flow matching for audio generation from omni-modalities},
  author={Xu, Xuenan and Mei, Jiahao and Zheng, Zihao and Tao, Ye and Xie, Zeyu and Zhang, Yaoyun and Liu, Haohe and Wu, Yuning and Yan, Ming and Wu, Wen and others},
  journal={arXiv preprint arXiv:2509.24391},
  year={2025}
}

@article{unimoeaudio,
    title={UniMoE-Audio: Unified Speech and Music Generation with Dynamic-Capacity MoE},
  author={Liu, Zhenyu and Li, Yunxin and Zhang, Xuanyu and Teng, Qixun and Jiang, Shenyuan and Chen, Xinyu and Shi, Haoyuan and Li, Jinchao and Wang, Qi and Chen, Haolan and others},
  journal={arXiv preprint arXiv:2510.13344},
  year={2025}
}

@article{mot,
    title={Mixture-of-transformers: A sparse and scalable architecture for multi-modal foundation models},
  author={Liang, Weixin and Yu, Lili and Luo, Liang and Iyer, Srinivasan and Dong, Ning and Zhou, Chunting and Ghosh, Gargi and Lewis, Mike and Yih, Wen-tau and Zettlemoyer, Luke and others},
  journal={arXiv preprint arXiv:2411.04996},
  year={2024}
}

@article{vlmo,
    title={Vlmo: Unified vision-language pre-training with mixture-of-modality-experts},
  author={Bao, Hangbo and Wang, Wenhui and Dong, Li and Liu, Qiang and Mohammed, Owais Khan and Aggarwal, Kriti and Som, Subhojit and Piao, Songhao and Wei, Furu},
  journal={Advances in neural information processing systems},
  volume={35},
  pages={32897--32912},
  year={2022}
}

@inproceedings{beit3,
    title={Image as a foreign language: Beit pretraining for vision and vision-language tasks},
  author={Wang, Wenhui and Bao, Hangbo and Dong, Li and Bjorck, Johan and Peng, Zhiliang and Liu, Qiang and Aggarwal, Kriti and Mohammed, Owais Khan and Singhal, Saksham and Som, Subhojit and others},
  booktitle={Proceedings of the IEEE/CVF Conference on Computer Vision and Pattern Recognition},
  pages={19175--19186},
  year={2023}
}

@article{roformer,
    title={Roformer: Enhanced transformer with rotary position embedding},
  author={Su, Jianlin and Ahmed, Murtadha and Lu, Yu and Pan, Shengfeng and Bo, Wen and Liu, Yunfeng},
  journal={Neurocomputing},
  volume={568},
  pages={127063},
  year={2024},
  publisher={Elsevier}
}

@article{vae,
    title={Auto-encoding variational bayes},
  author={Kingma, Diederik P and Welling, Max},
  journal={arXiv preprint arXiv:1312.6114},
  year={2013}
}

@article{qwen3,
    title={Qwen3 technical report},
  author={Yang, An and Li, Anfeng and Yang, Baosong and Zhang, Beichen and Hui, Binyuan and Zheng, Bo and Yu, Bowen and Gao, Chang and Huang, Chengen and Lv, Chenxu and others},
  journal={arXiv preprint arXiv:2505.09388},
  year={2025}
}

@article{libritts,
    title={Libritts: A corpus derived from librispeech for text-to-speech},
  author={Zen, Heiga and Dang, Viet and Clark, Rob and Zhang, Yu and Weiss, Ron J and Jia, Ye and Chen, Zhifeng and Wu, Yonghui},
  journal={arXiv preprint arXiv:1904.02882},
  year={2019}
}

@inproceedings{emilia,
    title={Emilia: An extensive, multilingual, and diverse speech dataset for large-scale speech generation},
  author={He, Haorui and Shang, Zengqiang and Wang, Chaoren and Li, Xuyuan and Gu, Yicheng and Hua, Hua and Liu, Liwei and Yang, Chen and Li, Jiaqi and Shi, Peiyang and others},
  booktitle={2024 IEEE Spoken Language Technology Workshop (SLT)},
  pages={885--890},
  year={2024},
  organization={IEEE}
}

@inproceedings{vggsound,
    title={Vggsound: A large-scale audio-visual dataset},
  author={Chen, Honglie and Xie, Weidi and Vedaldi, Andrea and Zisserman, Andrew},
  booktitle={ICASSP 2020-2020 IEEE International Conference on Acoustics, Speech and Signal Processing (ICASSP)},
  pages={721--725},
  year={2020},
  organization={IEEE}
}

@inproceedings{audiocaps,
    title={Audiocaps: Generating captions for audios in the wild},
  author={Kim, Chris Dongjoo and Kim, Byeongchang and Lee, Hyunmin and Kim, Gunhee},
  booktitle={Proceedings of the 2019 Conference of the North American Chapter of the Association for Computational Linguistics: Human Language Technologies, Volume 1 (Long and Short Papers)},
  pages={119--132},
  year={2019}
}

@article{wavcaps,
    title={Wavcaps: A chatgpt-assisted weakly-labelled audio captioning dataset for audio-language multimodal research},
  author={Mei, Xinhao and Meng, Chutong and Liu, Haohe and Kong, Qiuqiang and Ko, Tom and Zhao, Chengqi and Plumbley, Mark D and Zou, Yuexian and Wang, Wenwu},
  journal={IEEE/ACM Transactions on Audio, Speech, and Language Processing},
  volume={32},
  pages={3339--3354},
  year={2024},
  publisher={IEEE}
}

@article{fma,
    title={FMA: A dataset for music analysis},
  author={Defferrard, Micha{\"e}l and Benzi, Kirell and Vandergheynst, Pierre and Bresson, Xavier},
  journal={arXiv preprint arXiv:1612.01840},
  year={2016}
}

@inproceedings{mtgjamendo,
    title={The mtg-jamendo dataset for automatic music tagging},
  author={Bogdanov, Dmitry and Won, Minz and Tovstogan, Philip and Porter, Alastair and Serra, Xavier},
  booktitle={Machine learning for music discovery workshop, international conference on machine learning (ICML 2019)},
  pages={1--3},
  year={2019},
  organization={Long Beach, CA, United States}
}

@article{seedtts,
    title={Seed-tts: A family of high-quality versatile speech generation models},
  author={Anastassiou, Philip and Chen, Jiawei and Chen, Jitong and Chen, Yuanzhe and Chen, Zhuo and Chen, Ziyi and Cong, Jian and Deng, Lelai and Ding, Chuang and Gao, Lu and others},
  journal={arXiv preprint arXiv:2406.02430},
  year={2024}
}

@article{songdescriber,
    title={The song describer dataset: a corpus of audio captions for music-and-language evaluation},
  author={Manco, Ilaria and Weck, Benno and Doh, Seungheon and Won, Minz and Zhang, Yixiao and Bogdanov, Dmitry and Wu, Yusong and Chen, Ke and Tovstogan, Philip and Benetos, Emmanouil and others},
  journal={arXiv preprint arXiv:2311.10057},
  year={2023}
}

@inproceedings{whisper,
    title={Robust speech recognition via large-scale weak supervision},
  author={Radford, Alec and Kim, Jong Wook and Xu, Tao and Brockman, Greg and McLeavey, Christine and Sutskever, Ilya},
  booktitle={International conference on machine learning},
  pages={28492--28518},
  year={2023},
  organization={PMLR}
}

@article{wavlm,
    title={Wavlm: Large-scale self-supervised pre-training for full stack speech processing},
  author={Chen, Sanyuan and Wang, Chengyi and Chen, Zhengyang and Wu, Yu and Liu, Shujie and Chen, Zhuo and Li, Jinyu and Kanda, Naoyuki and Yoshioka, Takuya and Xiao, Xiong and others},
  journal={IEEE Journal of Selected Topics in Signal Processing},
  volume={16},
  number={6},
  pages={1505--1518},
  year={2022},
  publisher={IEEE}
}

@inproceedings{vggish,
    title={CNN architectures for large-scale audio classification},
  author={Hershey, Shawn and Chaudhuri, Sourish and Ellis, Daniel PW and Gemmeke, Jort F and Jansen, Aren and Moore, R Channing and Plakal, Manoj and Platt, Devin and Saurous, Rif A and Seybold, Bryan and others},
  booktitle={2017 ieee international conference on acoustics, speech and signal processing (icassp)},
  pages={131--135},
  year={2017},
  organization={IEEE}
}

@inproceedings{clap,
    title={Large-scale contrastive language-audio pretraining with feature fusion and keyword-to-caption augmentation},
  author={Wu, Yusong and Chen, Ke and Zhang, Tianyu and Hui, Yuchen and Berg-Kirkpatrick, Taylor and Dubnov, Shlomo},
  booktitle={ICASSP 2023-2023 IEEE International Conference on Acoustics, Speech and Signal Processing (ICASSP)},
  pages={1--5},
  year={2023},
  organization={IEEE}
}

@article{fad,
    title={Fr\'{e}chet audio distance: A metric for evaluating music enhancement algorithms},
  author={Kilgour, Kevin and Zuluaga, Mauricio and Roblek, Dominik and Sharifi, Matthew},
  journal={arXiv preprint arXiv:1812.08466},
  year={2018}
}

@inproceedings{f5tts,
    title={F5-tts: A fairytaler that fakes fluent and faithful speech with flow matching},
  author={Chen, Yushen and Niu, Zhikang and Ma, Ziyang and Deng, Keqi and Wang, Chunhui and JianZhao, JianZhao and Yu, Kai and Chen, Xie},
  booktitle={Proceedings of the 63rd Annual Meeting of the Association for Computational Linguistics (Volume 1: Long Papers)},
  pages={6255--6271},
  year={2025}
}

@article{tangoflux,
    title={Tangoflux: Super fast and faithful text to audio generation with flow matching and clap-ranked preference optimization},
  author={Hung, Chia-Yu and Majumder, Navonil and Kong, Zhifeng and Mehrish, Ambuj and Bagherzadeh, Amir Ali and Li, Chuan and Valle, Rafael and Catanzaro, Bryan and Poria, Soujanya},
  journal={arXiv preprint arXiv:2412.21037},
  year={2024}
}

@article{cosyvoice3,
    title={Cosyvoice 3: Towards in-the-wild speech generation via scaling-up and post-training},
  author={Du, Zhihao and Gao, Changfeng and Wang, Yuxuan and Yu, Fan and Zhao, Tianyu and Wang, Hao and Lv, Xiang and Wang, Hui and Ni, Chongjia and Shi, Xian and others},
  journal={arXiv preprint arXiv:2505.17589},
  year={2025}
}

@article{mingomni,
  title={Ming-omni: A unified multimodal model for perception and generation},
  author={AI, Inclusion and Gong, Biao and Zou, Cheng and Zheng, Chuanyang and Zhou, Chunluan and Yan, Canxiang and Jin, Chunxiang and Shen, Chunjie and Zheng, Dandan and Wang, Fudong and others},
  journal={arXiv preprint arXiv:2506.09344},
  year={2025}
}

@article{kimiaudio,
    title={Kimi-audio technical report},
  author={Ding, Ding and Ju, Zeqian and Leng, Yichong and Liu, Songxiang and Liu, Tong and Shang, Zeyu and Shen, Kai and Song, Wei and Tan, Xu and Tang, Heyi and others},
  journal={arXiv preprint arXiv:2504.18425},
  year={2025}
}

@article{qwen25omni,
  title={Qwen2.5-Omni Technical Report},
  author={Xu, Jin and Guo, Zhifang and He, Jinzheng and Hu, Hangrui and He, Ting and Bai, Shuai and Chen, Keqin and Wang, Jialin and Fan, Yang and Dang, Kai and Zhang, Bin and Wang, Xiong and Chu, Yunfei and Lin, Junyang},
  journal={arXiv preprint arXiv:2503.20215},
  year={2025}
}

@article{mimoaudio,
  title={MiMo-Audio: Audio Language Models are Few-Shot Learners},
  author={Zhang, Dong and Wang, Gang and Xue, Jinlong and Fang, Kai and Zhao, Liang and Ma, Rui and Ren, Shuhuai and Liu, Shuo and Guo, Tao and Zhuang, Weiji and others},
  journal={arXiv preprint arXiv:2512.23808},
  year={2025}
}

@article{funaudiochat,
  title={Fun-Audio-Chat Technical Report},
  author={Team, Tongyi Fun and Chen, Qian and Cheng, Luyao and Deng, Chong and Li, Xiangang and Liu, Jiaqing and Tan, Chao-Hong and Wang, Wen and Xu, Junhao and Ye, Jieping and others},
  journal={arXiv preprint arXiv:2512.20156},
  year={2025}
}
}

\newpage
\appendix
\section{Additional Implementation Details}
\label{sec:appendix-impl}

\subsection{Continuous Audio VAE}

\textbf{Architecture.}
Our autoencoder is a CNN--GAN of the \textsc{Stable Audio Open}~\citep{stableaudioopen}/\textsc{DAC}~\citep{dac} family with three deviations: (i) an iSTFT synthesis head replaces the time-domain transposed-convolution stack at the decoder output, (ii) self-attention layers are inserted at the lowest-resolution stages of both the encoder and the decoder, and (iii) a learned patch-with-\texttt{[CLS]} aggregator at the bottleneck further reduces the latent rate~\citep{mingomni}.

\textbf{Encoder and bottleneck.}
The encoder takes $44.1$\,kHz stereo waveforms through a stack of strided $1$D residual blocks (Snake activations, weight normalisation) whose cumulative stride of $2048\times$ yields a $21.5$\,Hz feature sequence; self-attention blocks at the lowest-resolution stages let the encoder reach beyond its convolutional receptive field. This sequence is right-padded to a multiple of $P{=}2$, split into patches of $P$ frames each augmented with a learnable \texttt{[CLS]} token, and processed by a small self-attention aggregator that mixes within and across patches. The \texttt{[CLS]} positions are collected and projected to the parameters of a diagonal Gaussian posterior $q(\mathbf{z}\!\mid\!\mathbf{x})=\mathcal{N}(\mu,\mathrm{diag}(\sigma^2))$ at $21.5/P\approx 10.75$\,Hz, so an AR-Flow block of $B{=}1.0$\,s corresponds to $11$ latent tokens.

\textbf{Decoder.}
The decoder linearly interpolates the latent back to $21.5$\,Hz, refines it through $1$D ResBlocks with symmetric self-attention insertions, and synthesises the waveform with an iSTFT head whose analysis grid matches the feature rate (hop $=2048$ samples, $n_{\text{fft}}=8192$). The head emits $2\cdot 2\cdot (n_{\text{fft}}/2{+}1)$ channels for two stereo-channel complex spectrograms, which a per-channel inverse short-time Fourier transform (Hann window) maps to the $44.1$\,kHz stereo output. Routing the final $2048\times$ of upsampling through a closed-form filterbank rather than through additional transposed convolutions removes the dominant source of phase artefacts in DAC-style decoders.

\textbf{Training.}
The VAE is trained on the corpora of Section~\ref{sec:appendix-data} at $44.1$\,kHz stereo with a multi-resolution STFT reconstruction loss, a KL term on the Gaussian posterior with free-bits regularisation, and an adversarial loss combining a multi-period time-domain discriminator with a multi-resolution complex-spectrogram discriminator (topology following the public \textsc{Stable Audio Open} release). We then freeze the VAE for the remainder of AudioCALM training.

\subsection{Training Corpora and Preprocessing}
\label{sec:appendix-data}

The training mixture covers three modalities: speech (\textsc{LibriTTS}~\citep{libritts} and the English subset of \textsc{Emilia}~\citep{emilia}), general sound (the audio of \textsc{VGGSound}~\citep{vggsound}, \textsc{AudioCaps}~\citep{audiocaps}, and \textsc{WavCaps}~\citep{wavcaps}), and music (\textsc{FMA}~\citep{fma} and \textsc{MTG-Jamendo}~\citep{mtgjamendo}). Only the official training splits are used; \textsc{LibriTTS} \emph{test-clean}, the \textsc{AudioCaps} and \textsc{Song-Describer}~\citep{songdescriber} evaluation sets, and \textsc{SeedTTS-eval}~\citep{seedtts} are held out to rule out test-set contamination. All audio is resampled to $44.1$\,kHz and converted to two-channel stereo, with mono sources upmixed by replicating the single channel.

\subsection{Description-Style Conditioning Pipeline}
\label{sec:appendix-conditioning}

This subsection details the offline pipeline that produces the dual textual conditions consumed by the body model (Section~\ref{sec:asymmetry}). For every training clip we store a \emph{short} variant---the verbatim transcript for speech, the original short caption for sound and music---and a \emph{long-form} variant generated by an audio-conditioned MLLM (\textsc{Gemini~3~Pro}). The two variants are sampled with equal probability per training step.

\textbf{Prompt design.}
We use a two-tier prompt: a shared \emph{system message} fixes the output format and the audio-grounding contract; a modality-specific \emph{user message} lists the attributes to surface, with explicit priority ordering, and (for speech) pins the transcript-delimiter rule. The two-tier split is what makes the long-form descriptions homogeneous in style across speech, sound, and music while still soliciting modality-relevant content. The templates we issue to \textsc{Gemini~3~Pro} are reproduced verbatim below.

\paragraph{System message (identical across modalities).}
\begin{quote}\small\ttfamily
You are an audio analyst. Listen to the supplied audio clip and write a single fluent paragraph that will be used as a text prompt for an audio-generation model---so describe what the audio \emph{sounds like}, not what is happening behind the recording.
\\[2pt]
Hard rules:
\\
(1) Every adjective and attribute must be supported by the audio itself. If you are not confident from the audio alone, \emph{omit} the attribute; do not guess and do not pad with generic adjectives.
\\
(2) Do not begin with ``This is\ldots'' or ``The audio shows\ldots''. Start directly with the most distinctive sonic content.
\\
(3) Plain prose only: no bullet points, no markdown headings, no bracketed metadata, no numerical confidence scores.
\\
(4) Single paragraph, $\le 200$ words.
\end{quote}

\paragraph{User message --- speech.}
\begin{quote}\small\ttfamily
Modality: speech. Surface, in roughly this order of priority: (i) speaker timbre, age range, and apparent gender; (ii) prosody---speaking rate, pitch range, emphasis pattern, emotional tone; (iii) recording acoustics---reverberation level, background-noise level, overall fidelity (e.g., studio-clean vs.\ mobile-recorded).
\\[2pt]
At the end of the paragraph, splice the verbatim transcript wrapped in delimiters \texttt{<spoken>}\,\ldots\,\texttt{</spoken>}. The text inside the delimiters must reproduce the transcript character-for-character; do not rephrase, expand, shorten, or translate it. The delimiter pair must appear exactly once and must not be nested.
\\[2pt]
Transcript: \{short\_variant\}
\end{quote}

\paragraph{User message --- general sound.}
\begin{quote}\small\ttfamily
Modality: general sound. Surface, in roughly this order of priority: (i) the dominant acoustic events and their likely physical sources; (ii) the scene or environment that ties the events together; (iii) the foreground/background structure of the soundscape; (iv) the temporal arc---which events come first, which recur, and how the energy evolves.
\\[2pt]
Do not invent specific named entities (particular cities, brands, individuals) that the audio cannot disambiguate; describe categories instead (``a busy city street'' rather than ``downtown Tokyo'').
\\[2pt]
Original short caption: \{short\_variant\}
\end{quote}

\paragraph{User message --- music.}
\begin{quote}\small\ttfamily
Modality: music. Surface, in roughly this order of priority: (i) genre or stylistic register and the apparent instrumentation; (ii) tempo, rhythmic feel, and overall mood; (iii) structural cues---recurring melodic or rhythmic motifs, dynamic arc, vocal vs.\ instrumental balance; (iv) production character---production cleanliness, stereo image, lo-fi/hi-fi flavor.
\\[2pt]
Do not specify exact BPM values, named musical keys, or named artists/albums even if you think you recognize them.
\\[2pt]
Original short caption: \{short\_variant\}
\end{quote}

\noindent The system message anchors a uniform prose style and the audio-grounding contract; the user messages enumerate attributes by priority, ban modality-specific failure modes (named entities for sound, BPM/keys/artists for music), and pin the speech delimiter convention. The audio-grounding clause is what makes an audio-conditioned MLLM strictly required: a text-only LLM cannot recover speaker timbre, recording acoustics, or musical mood from the bare transcript or short caption alone.

\textbf{Speech transcript delimiter.}
For speech, the prompt additionally instructs the MLLM to splice the verbatim transcript inside the long-form description, wrapped in explicit content delimiters \texttt{<spoken>}\,\ldots\,\texttt{</spoken>}. The body LM can therefore attend to a globally diffuse description and still see the exact spoken content for local content alignment, without the description-style and transcript-style conditioning collapsing into two separate prompts. The same delimiter convention is reused at inference for TTS and voice-cloning prompts.

To prevent the byte-pair tokenizer from fragmenting these markers into multiple sub-pieces (which would scatter the boundary signal across $\sim$$3$--$4$ tokens per delimiter and force the LM to re-learn the same opening/closing semantics from many distinct token combinations), both \texttt{<spoken>} and \texttt{</spoken>} are added as \emph{special tokens} to the body LM's tokenizer, extending the Qwen3 vocabulary by two entries. Each delimiter is then a single token at training and inference, and the LM has a single dedicated embedding row per boundary that it can attend to as a unit. The two new embedding rows are initialized as the mean of Qwen3's existing chat-template special tokens, placing them within the boundary-marker subspace the backbone already uses, and are trained jointly with the rest of the model.

\textbf{Quality control.}
For every captioning batch, $5\%$ of the produced long-form descriptions are sampled uniformly at random and reviewed by trained annotators. A description is rejected if it (i) asserts audio attributes the clip does not actually exhibit, (ii) omits attributes the prompt template asks for, or (iii) violates the speech delimiter convention. Rejected clips are returned to the MLLM for re-captioning under the same prompt; a clip that fails review more than three times is dropped from the training set entirely. Clips that pass review (and the unsampled remainder of their batch) enter the training manifest.

\textbf{Storage and loader behavior.}
Both variants are stored as parallel fields in the training manifest. At every step the data loader draws a Bernoulli($0.5$) coin per clip to select which variant is used as the conditioning text; the audio target is identical in both cases. This dual exposure is what teaches the LM to handle both terse prompts (a bare label) and richly described prompts at inference within a single conditioning interface.

\subsection{Hyperparameters and Training Setup}
\label{sec:appendix-hparams}

Table~\ref{tab:appendix-hparams} summarizes the architectural, training, regularization, and inference hyperparameters of the AudioCALM run that produces all results in Tables~\ref{tab:tts}, \ref{tab:t2a-music}, and \ref{tab:ablation}. The appendix sweeps (Tables~\ref{tab:appendix-backbone-scale}--\ref{tab:appendix-block-sweep}) differ only along the swept axis. All values were fixed before any evaluation on held-out splits: optimizer settings follow the Qwen3 pretraining recipe, the rectified-flow schedule and CFG dropout follow~\citep{sd3,rectifiedflow}, and the AR-Flow block size, exposure-bias scale, and clean-prefix noise scale were selected on a $5\%$ in-domain validation slice held out from the training mixture, with no held-out evaluation prompts used for selection.

\begin{table}[htbp]
\centering
\small
\vspace{-2mm}
\caption{Model, training, and inference hyperparameters for AudioCALM.}

\label{tab:appendix-hparams}
\begin{tabular}{p{0.45\linewidth} p{0.45\linewidth}}
\toprule
\textbf{Backbone and head} & \textbf{Value} \\
\midrule
Backbone init & Qwen3-1.7B \\
Hidden size / layers / heads & $2048$ / $28$ / $16$ (Qwen3 default) \\
A-MoME speech FFN width & matches shared FFN; zero-init \\
Flow-matching head $\phi_{\text{out}}$ & linear, zero-init \\
Stop head $g_\text{stop}$ & linear $\to$ sigmoid; ramp window $K_{\text{stop}} = 10$ \\
Audio embedding $\phi_{\text{in}}$ & linear; $C{=}64$ VAE channels $\to H{=}2048$ \\
Position embedding & shared RoPE with text \\
Timestep embedding $\tau(t)$ & sinusoidal, summed into $e_i^{(t)}$ \\
\midrule
\textbf{Training} & \\
\midrule
Optimizer & AdamW, $\beta_1{=}0.9$, $\beta_2{=}0.95$, weight decay $0$ \\
Learning rate & $1{\times}10^{-4}$ constant after $2{,}000$-step linear warmup \\
Gradient clipping & global norm $1.0$ \\
Total steps & $300{,}000$ \\
Global batch & $\sim\!64$\,k tokens ($8$\,k tokens / GPU, $8$ GPUs; token count includes training-time noise tokens) \\
Precision & bfloat16 mixed precision; FSDP HYBRID-SHARD \\
Sequence packing length & $4096$ tokens (text $+$ audio) \\
Modality mixing ratio (sp.\ /\ sn.\ /\ mu.) & $0.4$ / $0.3$ / $0.3$ at the example level \\
Caption mode (short / long) & $0.5$ / $0.5$ per example \\
Caption (CFG) dropout & $10\%$ unconditional during training \\
Rectified-flow timestep $t$ & $\sigma(u),\ u \sim \mathcal{N}(0,1)$ (logit-normal) \\
Per-block clean-prefix noise $\sigma_\text{clean}$ & $0.1$ \\
Exposure-bias perturbation $\gamma_\text{exp}$ & $0.1$ \\
Stop-head ramp length $K_{\text{stop}}$ & $10$ tokens \\
\midrule
\textbf{Inference} & \\
\midrule
AR-Flow block size $B$ & $1.0$\,s ($11$ latent tokens) \\
Rectified-flow steps per block $K_\text{flow}$ & $24$, Euler integration \\
CFG scale $w$ & $3.0$ \\
Stop probability threshold $\tau_\text{stop}$ & $0.95$ \\
Sampling temperature on velocity & $1.0$ (deterministic Euler) \\
\bottomrule
\end{tabular}
\vspace{-4mm}
\end{table}

\section{Evaluation Metrics}
\label{sec:appendix-metrics}

\subsection{Objective Metrics}

AudioCALM and all baselines are decoded at $44.1$\,kHz mono; audio is resampled \emph{per metric} to the rate expected by the underlying scoring model---$16$\,kHz for Whisper, WavLM-Large, and VGGish, and $48$\,kHz for LAION-CLAP---using a polyphase resampler with a Kaiser window.

\textbf{WER (speech).}
Intelligibility is measured as the word error rate of the generated speech when transcribed by the open-source \textsc{Whisper-large-v3}~\citep{whisper} ASR model under greedy decoding (\texttt{temperature}{=}$0$, \texttt{condition\_on\_previous\_text}{=}\texttt{False}, no language hint). Both the hypothesis and the reference target text pass through the standard Whisper text-normalization pipeline---Unicode \textsc{NFKC}, case folding, removal of punctuation and bracketed annotations, white-space collapse, and number-to-word expansion via the Whisper \texttt{EnglishTextNormalizer}---before computing a Levenshtein word distance with \texttt{jiwer}. WER is reported as a fraction in $[0,1]$; lower is better. We restrict speech evaluation to the \emph{en} split of SeedTTS-eval, so CER is not reported.

\textbf{SIM (speech).}
For zero-shot voice cloning we report SIM, the cosine similarity between the speaker embedding of the generated utterance and the speaker embedding of the prompt audio. Embeddings are extracted with \textsc{WavLM-Large}~\citep{wavlm} fine-tuned for speaker verification on VoxCeleb, using the model's mean-pooled embedding over the full utterance. On \textsc{LibriTTS}~\emph{test-clean} the prompt is a $3$\,s clip drawn from a \emph{different} utterance of the same speaker as the target sentence, following the protocol of~\citep{f5tts}; on \textsc{SeedTTS-eval}~(\emph{en}) we use the prompt audio and target text shipped with the benchmark verbatim. SIM lies in $[-1,1]$; higher is better.

\textbf{FAD (sound, music).}
Fr\'echet Audio Distance~\citep{fad} measures the distributional gap between generated and reference audio in the \textsc{VGGish}~\citep{vggish} embedding space. We use Google's reference \texttt{frechet\_audio\_distance} implementation: both candidate and reference clips are downmixed to $16$\,kHz mono and split into $0.96$\,s VGGish frames, per-clip embeddings are pooled into a single multivariate Gaussian per set, and the Fr\'echet distance between the two Gaussians is reported. The reference distribution is built from the ground-truth waveforms of the \textsc{AudioCaps} test split for sound and the \textsc{Song-Describer} evaluation pool for music; the candidate distribution contains exactly one generated clip per evaluation prompt at the same nominal duration as the corresponding reference. FAD is non-negative; lower is better.

\textbf{CLAP (sound, music).}
Text--audio alignment is reported as the cosine similarity between the LAION-\textsc{CLAP}~\citep{clap} text and audio embeddings of an evaluation caption and its corresponding generation. To allow direct comparison across sound and music we use a single checkpoint, \texttt{music\_audioset\_epoch\_15\_esc\_90.14}; the audio branch consumes $48$\,kHz mono input zero-padded or centre-cropped to the model's $10$\,s receptive field, while the text branch consumes the unmodified evaluation caption (no truncation is required at the caption lengths used). We report the per-prompt cosine similarity averaged over all prompts in the evaluation set; values lie in $[-1,1]$, higher is better.

\subsection{Subjective MOS Protocol}

For every system we additionally collect Mean Opinion Scores on a randomly sampled subset of $N{=}50$ prompts per benchmark. Speech is rated on a single \emph{overall-quality} scale, denoted \textbf{MOS} in Table~\ref{tab:tts}: because the target text is fixed and prompt adherence is already captured by WER, a separate text-relevance axis would be redundant. Sound and music are rated on two complementary axes: \textbf{MOS-Q} (overall perceived quality) and \textbf{MOS-T} (text relevance to the caption), the latter being essential in the absence of a ground-truth reference waveform. All axes use the standard $5$-point ITU-T~P.808 absolute-category-rating scale ($1$~bad, $2$~poor, $3$~fair, $4$~good, $5$~excellent), with anchor descriptions adapted per modality.

\textbf{Listening setup.}
Each rater completes the study in a single session of at most $30$\,min on a desktop or laptop browser; mobile devices are blocked. All clips are loudness-normalized to $-23$\,LUFS with a true-peak ceiling of $-1$\,dBTP following ITU-R~BS.1770-4 before being served, and playback level is fixed via a $1$\,kHz sine calibration trial at the start of the session. A short training stage shows three labelled reference clips per modality (anchored at the $1$, $3$, and $5$ levels of each axis) before the rating session begins, and three attention checks per session---an unmodified ground-truth clip, a $50\%$-time-stretched clip, and a clip with $20$\,dB additive white noise---are inserted at random positions. Sessions that fail more than one attention check are discarded and the affected ratings re-collected.

\subsection{Human-Subjects Study Operationalization}
\label{sec:appendix-mos-ops}

\textbf{Recruitment and demographics.}
Listeners were recruited through a vetted commercial crowdsourcing platform restricted to workers self-reporting fluent English and normal hearing. Each listener completed a one-minute pre-screening that included three suprathreshold tone-detection trials and a short comprehension question covering the rating instructions; only listeners who passed all four were admitted. The final pool comprised $\sim\!120$ unique listeners (anonymized to numeric IDs at intake), self-reporting a balanced gender distribution and an age range of $19$--$58$. No personally identifying information was collected, stored, or shared with the authors at any point.

\textbf{Consent and instructions.}
Before any audio was played, each listener was shown a plain-language consent screen describing the purpose of the study (``rating the perceived quality of computer-generated audio''), the data collected (numeric ratings and click metadata, no audio recordings of the listener), the right to withdraw at any time without forfeiting accumulated payment for completed clips, and a contact mailbox for questions or complaints. The full text of the consent screen, the rating instructions for each axis (MOS, MOS-Q, MOS-T) and modality (speech, sound, music), the calibration trial, and screenshots of the rating UI are included verbatim in the supplemental zip under \texttt{supp/mos\_protocol/}.

\textbf{Compensation.}
Listeners were paid a fixed task rate calibrated to a target hourly wage of \$$15$/hour, well above the federal and state minimum wage in both the data collector's and the platform's jurisdictions, and above the platform's recommended floor. Payment was issued for every completed session, including sessions whose ratings were later discarded for failing attention checks; only the affected ratings were re-collected from a different listener. Bonus payments were awarded for sessions with all three attention checks correct. Median completion time was $22$\,min, yielding a realized median rate of $\sim\!\$16.4$/hour after bonuses.

\textbf{Voice-clone consent in evaluation.}
Speaker-similarity (SIM) evaluation requires a reference voice prompt. We use only voices from the public \textsc{LibriTTS} \emph{test-clean} and \textsc{SeedTTS-eval}~(\emph{en}) splits, both of which were released by their authors with explicit re-distribution permissions for research; we do not record, solicit, or generate additional voice prompts from any human participant in this study, so no consent step beyond the upstream dataset releases is required for the SIM measurements.

\section{Additional Experimental Results}
\label{sec:appendix-exp}

\subsection{Backbone Scale Ablation}
\label{sec:appendix-backbone-scale}

We initialize AudioCALM from three Qwen3 checkpoints---0.6B, 1.7B (default), and 4B---training each for the same number of tokens as the main run. Table~\ref{tab:appendix-backbone-scale} reports per-modality WER (speech) and FAD/CLAP (sound, music).

\begin{table}[htbp]
\centering
\small
\caption{Backbone-scale ablation. All variants share the data, optimizer, and step budget of the main run. Bold marks the best result per column; the 1.7B row is the default used in the main results.}
\label{tab:appendix-backbone-scale}
\begin{tabular}{lccccc}
\toprule
Backbone & Sp.\ WER$\downarrow$ & Sn.\ FAD$\downarrow$ & Sn.\ CLAP$\uparrow$ & Mu.\ FAD$\downarrow$ & Mu.\ CLAP$\uparrow$ \\
\midrule
Qwen3-0.6B           & 0.025 & 2.38 & 0.31 & 3.75 & 0.31 \\
Qwen3-1.7B (default) & 0.020 & 1.95 & 0.37 & 2.02 & 0.36 \\
Qwen3-4B             & \textbf{0.018} & \textbf{1.89} & \textbf{0.39} & \textbf{1.97} & \textbf{0.38} \\
\bottomrule
\end{tabular}
\end{table}

\subsection{Classifier-Free Guidance Scale Sweep}
\label{sec:appendix-cfg-sweep}

We sweep the inference CFG scale $w \in \{1.0, 2.0, 3.0, 4.0, 5.0\}$ with all other hyperparameters fixed. 

\begin{table}[htbp]
\centering
\small
\caption{CFG scale sweep on the main evaluation benchmarks. Bold marks the best result per column; $w{=}3.0$ is the default used in the main results.}
\label{tab:appendix-cfg-sweep}
\begin{tabular}{cccccc}
\toprule
$w$ & Sp.\ WER$\downarrow$ & Sn.\ FAD$\downarrow$ & Sn.\ CLAP$\uparrow$ & Mu.\ FAD$\downarrow$ & Mu.\ CLAP$\uparrow$ \\
\midrule
1.0 & 0.028 & 3.22 & 0.31 & 2.49 & 0.32 \\
2.0 & 0.026 & 2.68 & 0.35 & 2.13 & 0.33 \\
\textbf{3.0} (default) & \textbf{0.020} & 1.95 & \textbf{0.37} & \textbf{2.02} & \textbf{0.36} \\
4.0 & 0.021 & \textbf{1.88} & 0.36 & 2.15 & 0.31 \\
5.0 & 0.020 & 2.01 & 0.32 & 2.30 & 0.30 \\
\bottomrule
\end{tabular}
\end{table}

\subsection{AR-Flow Block-Size Sweep}
\label{sec:appendix-block-sweep}

We sweep the AR-Flow block size $B \in \{0.25, 0.5, 1.0, 2.0, 4.0\}$\,s, holding the VAE frame rate and total decoded duration fixed, and additionally report a non-autoregressive baseline without block-wise decoding. Table~\ref{tab:appendix-block-sweep} reports per-modality quality. Smaller $B$ improves streaming granularity at the cost of slight quality regressions, while larger $B$ approaches the non-autoregressive regime.

\begin{table}[htbp]
\centering
\small
\caption{AR-Flow block-size sweep (in seconds). Bold marks the best result per column; $B{=}1.0$\,s is the default used in the main results.}
\label{tab:appendix-block-sweep}
\begin{tabular}{cccccc}
\toprule
$B$ (s) & Sp.\ WER$\downarrow$ & Sn.\ FAD$\downarrow$ & Sn.\ CLAP$\uparrow$ & Mu.\ FAD$\downarrow$ & Mu.\ CLAP$\uparrow$ \\
\midrule
0.25 & 0.019 & 2.45 & 0.35 & 2.48 & 0.33 \\
0.5  & \textbf{0.018} & 2.17 & 0.36 & 2.22 & 0.34 \\
\textbf{1.0} (default) & 0.020 & \textbf{1.95} & \textbf{0.37} & \textbf{2.02} & \textbf{0.36} \\
2.0  & 0.025 & 2.11 & 0.34 & 2.18 & 0.34 \\
4.0  & 0.031 & 2.08 & 0.33 & 2.14 & 0.33 \\
Non-Autoregressive  & 0.036 & 2.05 & 0.34 & 2.11 & 0.35 \\

\bottomrule
\end{tabular}
\end{table}

\subsection{Inference Step Sweep}
\label{sec:appendix-step-sweep}

\noindent
\begin{wrapfigure}[9]{r}{0.46\linewidth}
    \vspace{-22pt}
    \centering
    \includegraphics[width=\linewidth]{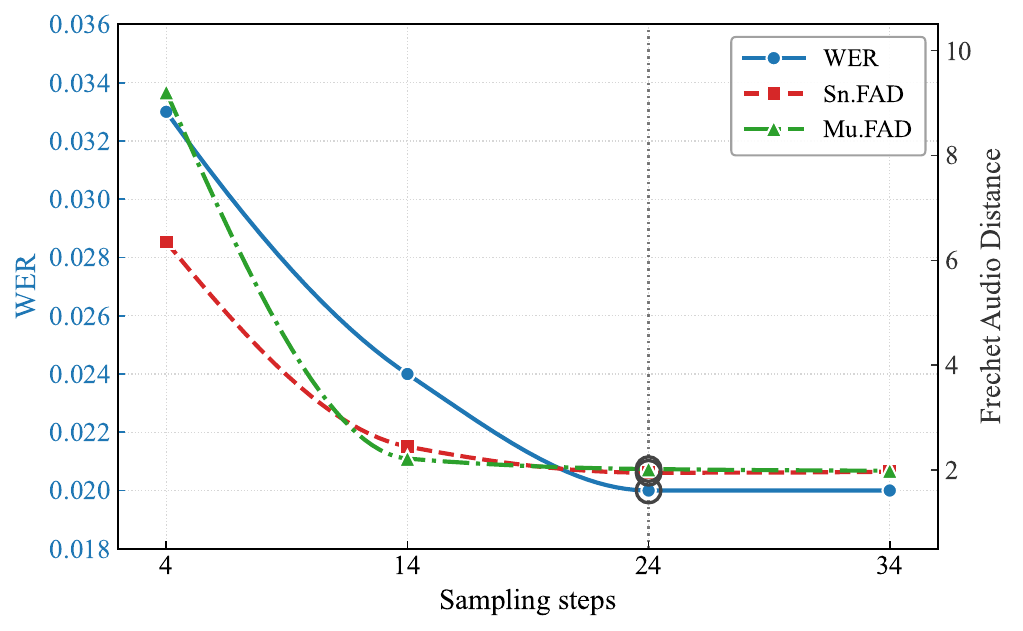}
    \caption{Inference-step sweep.}
    \label{fig:appendix-step-sweep}
    \vspace{-10pt}
\end{wrapfigure}

We sweep the number of inference steps with all other hyperparameters fixed. Figure~\ref{fig:appendix-step-sweep} shows that generation quality improves substantially as the number of inference steps increases, while the gains become marginal beyond roughly 20--30 steps. Based on this trade-off, we adopt 24 inference steps as the default configuration, which provides a favorable balance between generation quality and computational efficiency.

\section{Asset Licenses and Use}
\label{sec:appendix-licenses}

All training and evaluation assets used by AudioCALM are publicly released, open-source datasets, benchmarks, and pretrained models, cited at first use in Section~\ref{sec:appendix-data} (data) and Section~\ref{sec:appendix-metrics} (evaluators); we use each strictly within its open-source or research license. Our supplement releases only the AudioCALM weights, training and inference code, and the re-captioning prompts; no upstream dataset or pretrained checkpoint is re-hosted.

\section{Limitations}
\label{sec:appendix-limitations}

AudioCALM has three limitations. First, our training mixture is restricted to English speech and to publicly available sound and music corpora, so several domains---non-English speech, singing voice, and limited audio events---are not covered; in future work we plan to incorporate singing voice data so that the same backbone also supports song generation. Second, our backbone-scale study covers up to $4$\,B parameters (Section~\ref{sec:appendix-backbone-scale}); whether the trends we observe continue to hold at larger scales remains open and is left to future work. Third, although the VAE compresses audio to a relatively low frame rate, this work does not investigate long-form audio generation in depth, and we leave a focused study of long-horizon coherence and termination to future work.

\section{Broader Impacts}
\label{sec:appendix-broader}

A unified speech, sound, and music generator has positive applications in accessibility (text-to-speech for assistive devices), creative work (sound-design and music generation for independent creators), and research tooling (data augmentation for ASR and audio classification). The same capabilities raise misuse risks shared with the broader class of strong audio generators, most concretely zero-shot voice cloning and impersonation from a few seconds of reference audio, and the fabrication of plausible synthetic audio events that can be embedded in disinformation videos.

\section{Safeguards for Responsible Release}
\label{sec:appendix-safeguards}

We will release AudioCALM under a research-use license that explicitly prohibits non-consensual voice cloning, impersonation of identified individuals, and use in surveillance, harassment, or fraud workflows. The voice prompts used in our speaker-similarity evaluation come exclusively from public datasets whose original release permits such use (\textsc{LibriTTS}~\emph{test-clean} and \textsc{SeedTTS-eval}), and the MOS study (Section~\ref{sec:appendix-mos-ops}) collects no participant voice recordings, so this paper introduces no new voice-cloning targets. We do not claim that license clauses alone are sufficient against a determined adversary, and we regard parallel community work on synthetic-audio detection and content provenance as a necessary complement to this release.

\section{Declaration of LLM Usage}
\label{sec:appendix-llm-usage}

We use \textsc{Gemini~3~Pro} as the audio-conditioned captioner of the description-style conditioning pipeline (Section~\ref{sec:appendix-conditioning}), which contributes a measurable share of the gain over modality-specific baselines (ablation row~(c) of Table~\ref{tab:ablation}). This pass runs once, offline, on the training data only; no LLM is invoked in AudioCALM's inference path, and at evaluation the textual condition is the original benchmark text (verbatim transcripts for \textsc{LibriTTS}~\emph{test-clean} and \textsc{SeedTTS-eval}, original captions for \textsc{AudioCaps} and \textsc{Song-Describer}) rather than any Gemini-rewritten variant, so reproducing the reported numbers does not require querying \textsc{Gemini~3~Pro} at all. Because the captioner is closed-source, exact reproduction \emph{of training} without comparable multimodal-LLM access is not guaranteed; we mitigate this by releasing the captioning prompts verbatim and the cached annotations for every public training clip, so the rest of the pipeline can be reproduced without re-querying the captioner. Beyond this methodological role, the authors used a general-purpose writing assistant for copy editing of text already drafted by the authors; per the NeurIPS LLM policy, writing-only use does not require declaration and is mentioned here only for transparency.


\end{document}